\documentclass[prb,superscriptaddress,twocolumn,floatfix,showpacs]{revtex4-1}
\usepackage{graphicx,color}
\usepackage{amsmath,amssymb,bm}

\definecolor{darkred}{rgb}{0.90,0,0}
\definecolor{darkgreen}{rgb}{0,0.60,.2}
\definecolor{darkblue}{rgb}{0,0,1}
\definecolor{grey}{cmyk}{0,0,0,0.25}
\definecolor{orange}{cmyk}{0,0.6,0.8,0}

\begin{document}
\title{\boldmath Light scattering and dissipative dynamics of many fermionic atoms in an optical lattice}

\author{S.~Sarkar}
\affiliation{Department of Physics and Astronomy, University of Pittsburgh, Pittsburgh, Pennsylvania 15260, USA}
\author{S.~Langer}
\affiliation{Department of Physics and Astronomy, University of Pittsburgh, Pittsburgh, Pennsylvania 15260, USA}
\author{J.~Schachenmayer}
\affiliation{JILA, NIST, Department of Physics, University of Colorado, 440 UCB, Boulder, CO 80309, USA}
\affiliation{Department of Physics and Astronomy, University of Pittsburgh, Pittsburgh, Pennsylvania 15260, USA}
\author{A.~J.~Daley}
\affiliation{Department of Physics and Astronomy, University of Pittsburgh, Pittsburgh, Pennsylvania 15260, USA}
\affiliation{Department of Physics and SUPA, University of Strathclyde, Glasgow G4 0NG, Scotland, U.~K.}

%

\begin{abstract}

We investigate the many-body dissipative dynamics of fermionic atoms in an optical lattice in the presence of incoherent light scattering. Deriving and solving a master equation to describe this process microscopically for many particles, we observe contrasting behaviour in terms of the robustness against this type of heating for different many-body states. In particular, we find that the magnetic correlations exhibited by a two-component gas in the Mott insulating phase should be particularly robust against decoherence from light scattering, because the decoherence in the lowest band is suppressed by a larger factor than the timescales for effective superexchange interactions that drive coherent dynamics. Furthermore, the derived formalism naturally generalizes to analogous states with SU(N) symmetry. In contrast, for typical atomic and laser parameters, two-particle correlation functions describing bound dimers for strong attractive interactions exhibit superradiant effects due to the indistinguishability of off-resonant photons scattered by atoms in different internal states. This leads to rapid decay of correlations describing off-diagonal long-range order for these states. Our predictions should be directly measurable in ongoing experiments, providing a basis for characterising and controlling heating processes in quantum simulation with fermions.
\end{abstract}

\pacs{37.10.Jk, 42.50.-p, 67.85.Hj}
\maketitle

\section{Introduction}
\label{sec:intro}

In recent years, there has been remarkable progress towards quantitative applications of quantum simulators \cite{Cirac2012,Bloch2012,Blatt2012,Guzik2012} to the study of many-body physics in strongly interacting systems \cite{Bloch2006,Bloch2008}. In the case of fermionic atoms in optical lattices, the level of microscopic understanding and control achieved opens the door towards the study of physics associated with the Hubbard model \cite{Esslinger2010}, including magnetic correlations in two-component fermi gases in an optical lattice \cite{Greif2013}, and going beyond this to the study of many-body physics with high degrees of SU(N) symmetry in group-II-like atoms \cite{Cazalilla2009,Gorshkov2010,Scazza2014,Zhang2014,Cazalilla2014}. At the same time, these experiments provide an ideal environment for probing out-of-equilibrium physics \cite{Cazalilla2010,Cazalilla2011}, both in terms of quench dynamics \cite{Heidrich-Meisner2008,del-Campo2008,Heidrich-Meisner2009,Langer2012,Bolech2012,Schneider2012,Vidmar2013}, and also dissipative dynamics\cite{Kantian2009,Sandner2011,Yi2012,Bernier2013}. Controlled dissipative dynamics \cite{Muller2012} could be used in these systems to observe the emergence of pairing \cite{Yi2012} or counter-intuitive long-range correlations \cite{Bernier2013}, as well as Pauli-blocking and quantum Zeno effects in dissipative processes \cite{Sandner2011,Kantian2009}. 

A key challenge for state-of-the art experiments with multiple spin species of fermions \cite{Cirac2012,Bloch2012} is to realize entropies per particle low enough to observe quantum magnetism or other physics on small energy scales. This is particularly true of strongly interacting regimes $|U|\gg J$, where $J$ is the tunnelling amplitude to neighbouring sites, and $U$ is the on-site interaction energy, in which the dominant order is often driven by small terms \cite{Jordens2010,McKay2011,Hofstetter2002,Mathy2012,Jordens2008,Paiva2010,De-Leo2011,Fuchs2011} of the order of $J^2/U$. Characterization and control over heating therefore takes on a special importance. From a theoretical point of view such dynamics are complicated, as they involve understanding the interplay between few-particle atomic physics, and out-of-equilibrium many-body dynamics. However, the atomic physics of these systems is sufficiently well understood that we can look to derive microscopic models describing heating of many-particle systems, either with technical noise such as fluctuations of an optical lattice potential \cite{Pichler2012,Pichler2013}, or with incoherent light scattering for one atom \cite{Gordon1980,Dalibard1985}, up to many bosons \cite{Gerbier2010,Pichler2010,Poletti2012,Poletti2013,Schachenmayer2014}. For particles in optical lattices, incoherent light scattering gives a fundamental limit to coherence times of many-body states, and therefore minimal requirements for our understanding of out-of-equilibrium many-body dynamics in quantum simulators \cite{Jordens2010,McKay2011,Trotzky2010}.

Here we study in detail the dissipative dynamics of many fermionic atoms in a far-detuned optical lattice due to incoherent light scattering, finding effective decoherence rates for the many-body states that strongly depend on the details of the many-body state. In particular, we find contrasting results for magnetic order with strong repulsive interactions -- where the states are very robust, and for strong attractive interactions -- where the coherence of a gas of bound dimers is rapidly reduced. In order to obtain these results, we consider the atomic physics of group-I and group-II atoms, and derive a many-body master equation \cite{Gardiner2010,Gardiner1985,Lax1963,Collett1984} that provides a microscopic description, independent of the lattice geometry or dimensionality. Using analytical techniques and by combining time-dependent density matrix renormalisation group methods \cite{Vidal2004,Daley2004,White2004,Verstraete2008,Schollwock2011} with quantum trajectory techniques \cite{Daley2014,Carmichael1993,Molmer1993,Dum1992,Daley2009}, we solve the master equation and determine the decoherence dynamics of a range of initial many-body states. The resulting formalism can also be straight-forwardly generalised to heating of group-II atoms with SU(N) magnetic order, making the results we obtain here directly observable in and relevant for a large variety of ongoing experiments. 
\section{Overview}
\subsection{Summary of results for atoms in the lowest Bloch band} 
When we begin with atoms in the lowest Bloch band of the optical lattice, and consider the decoherence of many-body states, we observe strikingly contrasting results in different regimes. This is especially true for the case of strong interactions $|U| \gg J$. While the rate of spontaneous emissions $\gamma$ in typical current experiments can be of the order of 0.001 -- 0.01 $J$, the physics of magnetically ordered states with $U>0$ and bound dimers for $U<0$ is driven by terms that arise in second order perturbation theory as $\propto J^2/U$. Conservatively taking $U\gtrsim 10 J$ in this regime, we have $\gamma \gtrsim$ 0.01 -- 0.1 $J^2/U$. Therefore, there is a danger that these states may be particularly susceptible to decoherence at a rate that is relatively fast compared to the relevant dynamical timescales. 

We find this concern to be well founded in the case of attractive interactions $U<0$ with equal filling of two spin species. In that regime a gas of bound dimers exists, in which dimers tunnel in perturbation theory with amplitude $2J^2/U$, and forms a superfluid with long-range order at low energy. However, this order is strongly  susceptible to spontaneous emissions: correlations describing off-diagonal order of dimers decay at a rate not only given by $2\gamma$ (the rate of scattering for two independent particles), but instead $4\gamma$. This arises from additional superradiant enhancement due to the indistinguishability of off-resonant photons scattered by atoms in different internal states.

This is in strong contrast to the case of magnetic ordering for repulsive interactions $U>0$, where minimising interaction energy favours insulating states with a single atom per site, and magnetic ordering driven by a superexchange interaction of amplitude $2J^2/U$. We show that these states can be particularly robust, which can be intuitively understood as follows: for the typical experimental case, direct spin decoherence does not occur because the lattice lasers are far detuned and photons scattered by atoms in different internal states are indistinguishable. As a result, spontaneous emission only distinguishes between different on-site particle number states, and not different spin states. This suppresses decoherence by a factor related to the probability of doubly-occupied or unoccupied lattice sites (assuming we begin with unit filling), which in higher dimensions is of order $J^2/U^2$. As a result, the dominant process involves transfer of particles to higher Bloch bands, which itself is suppressed by the Lamb-Dicke factor $\eta \sim 0.1$. Hence, these magnetically ordered states should exhibit a particular robustness against decoherence due to spontaneous emissions. 

The general formalism derived here applies to higher dimensions and can also be straight-forwardly generalised to larger numbers of internal states, including nuclear spins in group-II atoms. In particular, the results we obtain in perturbation theory showing the robustness of magnetic order can be directly generalised to the heating of group-II atoms with SU(N) magnetic order. 

\subsection{Outline of this article}

This paper is organized as follows: In Sec.~\ref{Sec:atomic} we summarise the atomic physics of a single group-I atom or group-II atom and justify the microscopic assumptions we use as basis for describing the many-body dynamics. In Sec.~\ref{Sec:master} we outline the derivation of the many-body master equation for light scattering by fermionic atoms. Sec.~\ref{Sec:double} presents the intuitive regime of atoms in a double-well potential, in preparation for Section~\ref{Sec:numerics}, where we study the full many-body dynamics on a lattice. We present a summary and outlook in Sec.~\ref{Sec:sum}. More technical details of the calculations in Sec.~\ref{Sec:atomic} and Sec.~\ref{Sec:master} are organized in Appendix A and B respectively.

\section{Atomic physics}
\label{Sec:atomic} 

To provide the framework for the derivation of the master equation, we summarise the relevant atomic physics for the atomic species predominantly used in experiments with ultra-cold atoms. Group-I (alkali-metal) atoms have been used widely, and recently group-II (alkaline earth-metal) atoms have been established for the realisation of systems with SU(N) symmetry \cite{Scazza2014,Zhang2014} . We consider spontaneous emissions when an atom is trapped in an optical lattice created by far-detuned laser field and show that the atom returns to the same state it started from  with very high probability. In addition the scattered photons are also indistinguishable, resulting in low direct spin decoherence. In the following we provide two prototypical examples, $^{171}\text{Yb}$ (group-II) and $^6\text{Li}$ (group-I).

\subsection{Group-II atoms}

\begin{figure}[t]
\centering
\includegraphics[width=0.45\textwidth]{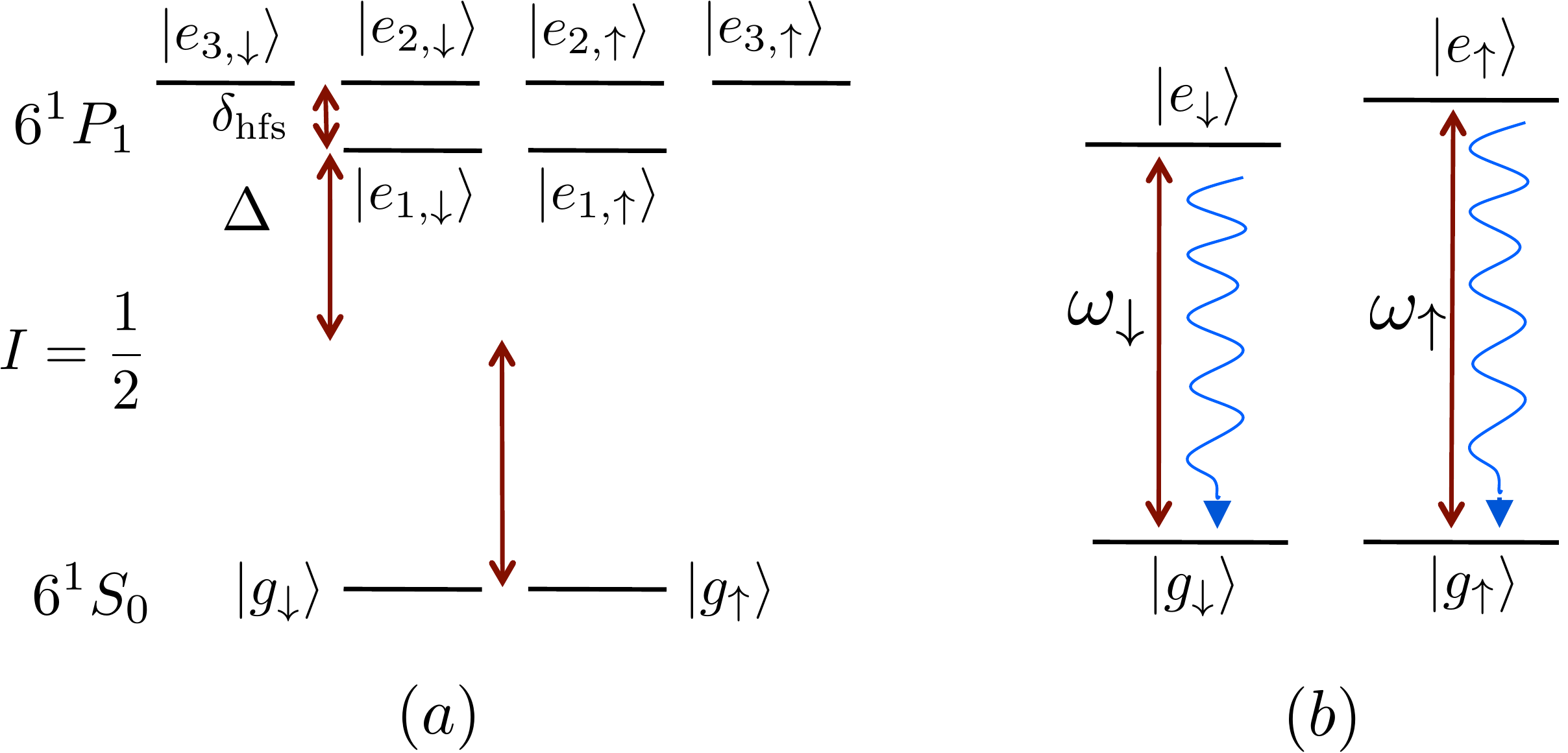}
\caption{(a) Atomic structure of $^{171}\text{Yb}$ (nuclear spin $I=1/2$). We use the spectroscopic notation for the sublevels and show hyperfine structure splittings of the lowest singlet levels (energies are not drawn to scale). The ground states have total electron spin of zero and states in this manifold essentially only differ in the nuclear spin component $m_I$. We write the two ground states as spin down and spin up states for $m_F=-1,1$ respectively. (b) Reduction of this hyperfine structure to an effective four-level system where, for very large detuning ($\Delta \gg \delta_{\text{hfs}}, |\omega_{\uparrow}-\omega_{\downarrow}|$), we can neglect the possibility of a spin flip and can take the photons scattered from each spin system to be identical. \label{Yb}}
\end{figure}

First we look at the case of group-II atoms, specifically $^{171}\text{Yb}$ (Fig.~\ref{Yb}(a)). These atoms with two valence electrons have a ground state which is a spin singlet, with zero total electronic angular momentum. Hence, the ground states differ only in the z-component of the nuclear spin, $I=1/2$, and we have two states in the lowest manifold. The electric field of the laser only couples directly to the orbital motion of the electron, and we can define a detuning $\Delta$ from the most closely coupled excited level, e.g., $^1$P$_1$ (using spectroscopic notation), as the difference between the laser frequency and atomic transition frequency. If the field is far detuned, i.e., $\Delta$ is large compared with the hyperfine structure energy splitting $\delta_{\text{hfs}}$, then the individual hyperfine states cannot be resolved, and the hyperfine coupling cannot be used to rotate the nuclear spin state during spontaneous emissions. Phrased in a different way, we can note that  a particular choice of ground state is always coupled to a superposition of excited hyperfine states, which depends on the detuning. For large detuning this superposition is such that when the atoms return to the ground state, decay channels corresponding to a spin flip interfere destructively and its relative rate is of the order $\sim (\delta_{\text{hfs}}/\Delta)^2$ (see Appendix \ref{sec:Atomic Physics} for an explicit calculation). 

In typical experimental setups where $\delta_{\text{hfs}}\approx 324$ MHz  \cite{Berends1992,Reichenbach2007}, and far-off-resonance lattices can be detuned by tens or hundreds of nanometers ($\sim 10^{14}$Hz), this rate of spin-flips is extremely small. In such a limit, the group-II atomic system can be regarded as an assembly of two decoupled two-level systems for the two different nuclear spin states (Fig.~\ref{Yb}(b)). Relative shifts of the transition frequencies between the levels are small, but to account for any small difference, we define transition frequencies $\omega_{\uparrow}$ and $\omega_{\downarrow}$, as shown in Fig.~\ref{Yb}(b). For $\Delta \gg |\omega_{\uparrow}-\omega_{\downarrow}|$, the relative frequencies of the scattered photons cannot be resolved \cite{Cohen-Tannoudji1992}, resulting in suppression of direct spin decoherence. We explore the differences between identical and non-identical photon scattering in more detail in Sec.~\ref{Sec:master}. 

\subsection{Group-I atoms}

\begin{figure}[t]
\centering
\includegraphics[width=0.3\textwidth]{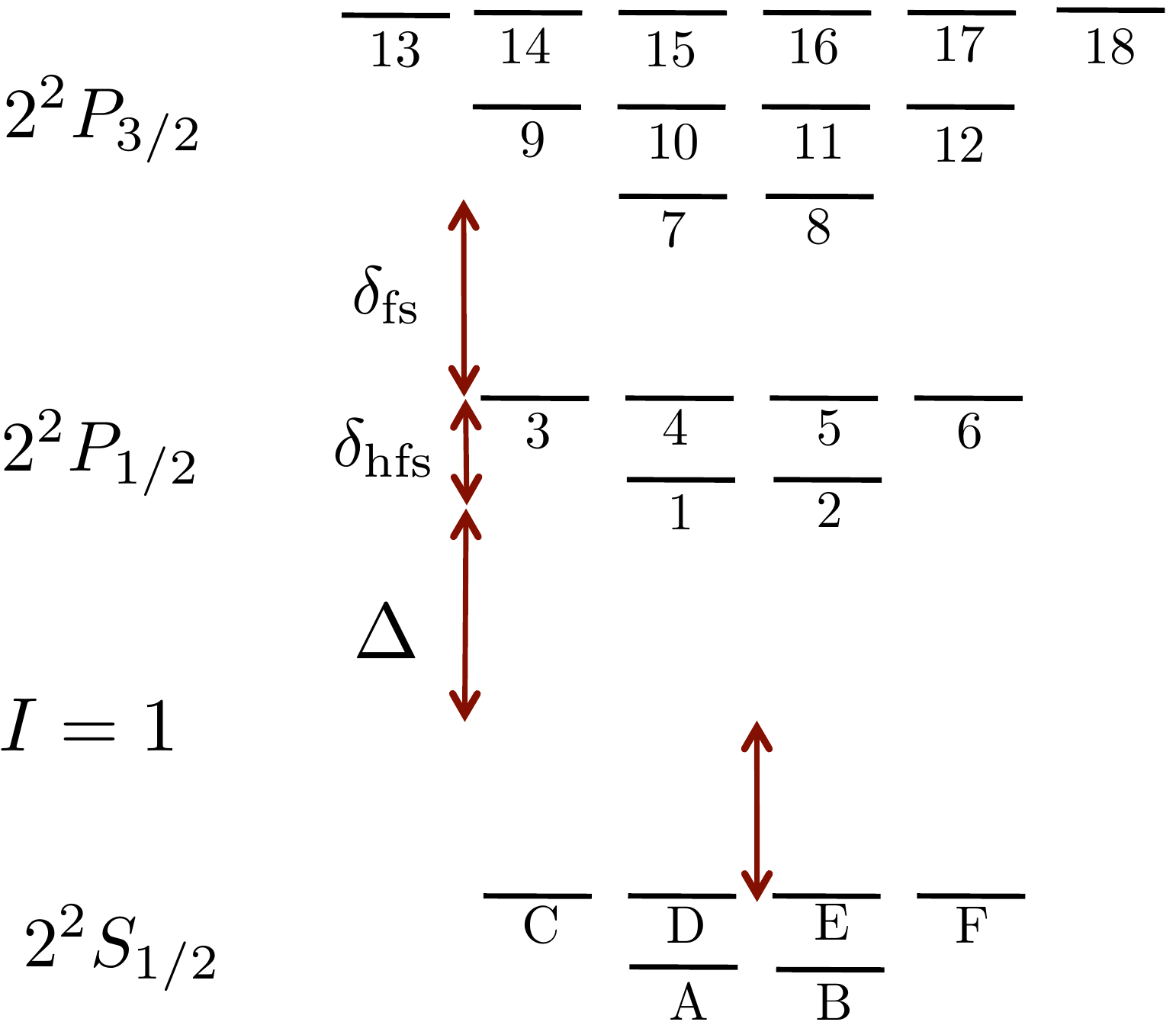}
\caption{A diagram of the atomic structure of $^{6}\text{Li}$ (nuclear spin $I=1$) showing the lowest hyperfine manifolds (energies not drawn to scale). The ground states in the $S$ sublevel are labeled by $A,B,\dots,F$ and the excited states in $P$ sublevels are labeled by $1,2,\dots,18$. These names will be used in the text in discussing transitions between different levels.\label{Li}}
\end{figure}

In addition to the considerations in the group-II case, group-I atoms such as $^6\text{Li}$ have nonzero electron spin in the ground state. We then need to consider the role of fine structure coupling and include excited levels in $^2P_{1/2}$ and $^2P_{3/2}$ (as shown in Fig.~\ref{Li}). These have a fine structure energy difference $\delta_{\text{fs}}$ between them and hyperfine structure energy splittings $\delta_{\text{hfs},P_{1/2}}$ and $\delta_{\text{hfs},P_{3/2}}$ within each manifold of states. Analogously to the group-II case, spin-flip processes that must change the nuclear spin are suppressed if the detuning is much larger than the hyperfine structure splitting, and also spin-flip processes changing the electronic spin are suppressed when the detuning is much larger than $\delta_{\text{fs}}$.  An example of a spin flip between two ground states that have different electron spins is $|g_D\rangle \rightarrow |g_E\rangle$ in Fig.~\ref{Li}. The relative rate of spin flip processes is $\propto (\delta_{\text{fs}}/\Delta)^2$. An example of the flip of a nuclear spin is $|g_D\rangle \rightarrow |g_A\rangle$, where the relative spin flip rate from a laser polarized along $z$-axis is $\propto (\delta_{\text{hfs},P_{1/2}}/\Delta-\delta_{\text{hfs},P_{3/2}}/\Delta)^2$, with constants that can be computed from the different dipole matrix elements. For a hyperfine structure splitting of $\delta_{\text{hfs}}=26.1$ MHz for $P_{1/2}$ and $4.5$ MHz for $P_{3/2}$, and $\delta_{\text{fs}}=10.05$ GHz for the fine structure splitting\cite{Libbrecht1995}, we can again assume that the spin-flip processes are very strongly suppressed in far-detuned lattices and are negligible on experimentally relevant timescales. More details of these calculations can be found in Appendix \ref{sec:Atomic Physics}.

This conclusion also holds well when we consider the role of an external magnetic field, including in the Paschen-Back regime.  For both $^{171}\text{Yb}$ and $^6\text{Li}$, the spin flip rates stay negligibly small even when an external magnetic field is introduced (see Appendix~\ref{sec:Atomic Physics}). For very high magnetic fields though the basic assumption  $\Delta \gg |\omega_{\uparrow}-\omega_{\downarrow}|$ has to be carefully revisited as the frequencies of the spontaneously emitted photons from different spin states are different now.

\section{Master Equation for Fermionic Many Body Systems}
\label{Sec:master}

We now derive the master equation describing fermionic atoms with two internal states, trapped in an optical lattice created by a far-detuned laser field and undergoing spontaneous emissions. We begin from the collective coupling of many atoms to the external radiation field (which we consider as the reservoir or bath), and obtain the equation of motion for the reduced atomic density operator for the motion of the atoms, $\rho$ (traced over the bath) in the form ($\hbar\equiv 1$)
\begin{align}
\frac{d}{dt}\rho=-i[H,\rho]+\mathcal{L}\rho
\,. \end{align}
Here the Hamiltonian $H$ describes the coherent dynamics whereas the Liouvillian $\mathcal{L}\rho$ corresponds to the dissipative dynamics due to spontaneous emission events \cite{Gardiner2010}. Note that this derivation is analogous to the case of a single species of bosonic atoms treated in Ref.~\onlinecite{Pichler2010}. Despite the different particle statistics and in the presence of an additional internal degree of freedom we remarkably obtain qualitatively equivalent terms and a very similar overall structure to the master equation. However, our generalization now takes into account the effects on the internal state dynamics for multiple electronic ground states, and below we will use this to investigate in detail the interplay between the motional dynamics and correlations in spin-ordered states. The master equation we derive is, however, very general, and can be applied directly to describe fermions in a rich variety of regimes in an optical lattice\cite{Lewenstein2007}. Furthermore, while we focus here on the two-species case, we see from the structure of our calculation that both the master equation and the conclusions for spin-ordered states can be straightforwardly generalized to SU(N) spin systems \cite{Cazalilla2014,Gorshkov2010}. 
 
In our treatment we take an ensemble of atoms, each with a mass $m$, and with four accessible internal states, electronic ground and excited states $|g\rangle$ and $|e\rangle$ for each of two spin manifolds, giving rise to the four-level systems depicted in Fig.~\ref{Yb}(b). Initially, we will consider the limit where $\omega_{\uparrow}=\omega_{\downarrow}=\omega_{eg}$, so that the photons emitted are indistinguishable between the different states. However, we will come back to check this assumption at the end of this section. The system is driven by a laser with frequency $\omega_L$ far detuned from the transition frequency by an amount $\Delta=\omega_L-\omega_{eg}$. Therefore the interactions between the atom and the laser light involve a spatially dependent Rabi frequency $\Omega(\bold{x})$ which is proportional to the laser field strength and to the dipole moment the atom, $d_{eg}$. To write down the master equation in second quantization we define the spin ($s$) dependent field operators $\psi_s(\bold{x})$ and they obey fermionic anti-commutation relations $\{\psi_s(\bold{x}),\psi^{\dag}_{s'}(\bold{y})\}=\delta_{s,s'}\delta(\bold{x}-\bold{y})$. In order to properly account for interactions, as well as losses from short-range contributions, we use standard arguments to separate the dominant contribution to the dynamics at large distances from the short-range physics \cite{Dalibard1998,Castin2001}. This gives rise to interaction terms which for a dilute gas at low scattering energies can be completely characterised by the s-wave scattering length, and for which losses, e.g., due to laser-assisted collisions at short distances can be accounted for via a small imaginary part of this length \cite{Fedichev1996}. The far detuned laser drive allows adiabatic elimination of the atoms in the excited states\cite{Lehmberg1970} and working in a frame rotating with the laser frequency we obtain a master equation of the form (see Appendix \ref{sec:Master Equation})

\begin{align}
\frac{d}{dt}\rho=-i\left(H_{\text{eff}}\rho-\rho H_{\text{eff}}^{\dag}\right)+\mathcal{J}\rho
\,. \label{ME} \end{align}
Here the non-hermitian effective Hamiltonian is:

\begin{align}
H_{\text{eff}}=H_0+H^{\text{light}}_{\text{eff}}+H^{\text{int}}_{\text{eff}}
\,.\end{align}
This effective Hamiltonian describes in addition to the coherent dynamics and the collisional processes also the scattering processes that transfer away the ground state population (therefore not trace preserving). The first term, $H_0$ is the Hamiltonian for non-interacting atoms in an optical lattice potential originating from the ac-Stark shift \cite{Pethick2008} induced by a standing wave of laser light:

\begin{align}
H_0=\sum_s\int d^3x\psi_s^\dag(\bold{x})\left(\frac{\nabla^2}{2m}+\frac{|\Omega(\bold{x})|^2}{4\Delta}\right)\psi_s(\bold{x}) 
\,.\end{align}

To model spontaneous emissions we couple the atoms to a radiation bath, namely the vacuum modes of the laser field. The effective Hamiltonian describing the atom-light interaction is given by:

\begin{widetext}
\begin{align}
H^{\text{light}}_{\text{eff}}=& \sum_{s,s'}\Gamma\int d^3xd^3yG(k_{eg}\bold{r})\frac{\Omega(\bold{y})\Omega^*(\bold{x})}{4\Delta^2}\psi_s^\dag(\bold{x})\psi_{s'}^\dag(\bold{y})\psi_{s'}(\bold{y})\psi_s(\bold{x}) \nonumber\\    
                   & -i\frac{\Gamma}{2}\sum_s\int d^3x\frac{|\Omega(\bold{x})|^2}{4\Delta^2}\psi_s^\dag(\bold{x})\psi_s(\bold{x})-i\frac{\Gamma}{2}\sum_{s,s'}\int d^3xd^3y\frac{\Omega(\bold{y})\Omega^*(\bold{x})}{4\Delta^2} F(k_{eg}\bold{r})\psi_s^\dag(\bold{x})\psi_{s'}^\dag(\bold{y})\psi_{s'}(\bold{y})\psi_s(\bold{x})
\,,\end{align}
\end{widetext}
where functions $F$ and $G$ are defined as 

\begin{align}
F(\boldsymbol{\xi})=& \frac{3}{2}\Bigg\{[1-(\boldsymbol{\hat{\xi}}\cdot\bold{\hat{d}}_{eg})^2]\frac{\sin{\xi}}{\xi} \nonumber \\
& +[1-3(\boldsymbol{\hat{\xi}}\cdot\bold{\hat{d}}_{eg})^2]\left(\frac{\cos{\xi}}{\xi^2}-\frac{\sin{\xi}}{\xi^3}\right) \Bigg\}
\,, \label{Ffn} \end{align}

\begin{align}
G(\boldsymbol{\xi})=& \frac{3}{4}\Bigg\{-[1-(\boldsymbol{\hat{\xi}}\cdot\bold{\hat{d}}_{eg})^2]\frac{\cos{\xi}}{\xi} \nonumber \\
& +[1-3(\boldsymbol{\hat{\xi}}\cdot\bold{\hat{d}}_{eg})^2]\left(\frac{\sin{\xi}}{\xi^2}+\frac{\cos{\xi}}{\xi^3}\right) \Bigg\}
\,,\end{align}
and $\Gamma$ is the Wigner-Weisskopf spontaneous decay rate.
The first term in $H^{\text{light}}_{\text{eff}}$ gives the dipole-dipole (created by photon exchange) interaction energy. The second term contains single-atom processes which absorb and then emit laser photons. The third term describes a collective two-atom excitation and de-excitation that can give rise to superradiance or subradiance in appropriate limits \cite{Lehmberg1970,Lehmberg1970a}. Now as $G$ decays as a function of inter-atomic distance we can focus only on interaction on a small scale set by the laser wavelength. At very short distances ($k_{eg}r\rightarrow 0$) it is possible to absorb the dipole-dipole interaction as a small modification to the collisional interactions, 
\begin{align}
H^{\text{int}}_{\text{eff}}=\int d^3x\,g(\bold{x})\psi_{\uparrow}^\dag(\bold{x})\psi_{\downarrow}^\dag(\bold{x})\psi_{\downarrow}(\bold{x})\psi_{\uparrow}(\bold{x})
\,.\end{align}
This term contains short range low-energy two-body scattering processes in the atomic system, characterized by a single parameter, the scattering length $a_s$. The same scattering length can be obtained using a pseudo-potential in $H^{\text{int}}_{\text{eff}}$ which is a contact potential \cite{Dalibard1998,Castin2001} with $g=4\pi\hbar^2a_s/m$. Now in the presence of laser light we also need to take into account light assisted collisional interactions. A red-detuned laser can give rise to optical Feshbach resonance resulting in modification of the scattering length which will depend on the laser intensity \cite{Chin2010,Fedichev1996}. This spatial dependence is reflected in $g(\bold{x})$ and away from the resonance we would get back $g \sim 4\pi\hbar^2a_s/m$. Loss of atoms due to light assisted collisions can be contained in an intensity dependent (and thus spatially dependent) imaginary part to the scattering length. As the rate of such losses are much less than the scattering rate, we can work in a regime where such loss processes have not occurred and can therefore leave out the imaginary part. Higher order corrections such as three-body collisions have also not been considered in this Hamiltonian as we work with dilute gases.

The last term in the master equation is the recycling term:

\begin{align}
\mathcal{J}\rho & =\Gamma\int d^3xd^3y\frac{\Omega(\bold{y})\Omega^*(\bold{x})}{4\Delta^2} F(k_{eg}\bold{r}) \nonumber \\
& \left(\sum_s\psi_s^\dag(\bold{y})\psi_s(\bold{y})\right)\rho\left(\sum_s\psi_s^\dag(\bold{x})\psi_s(\bold{x})\right)
\,,\end{align}
which contains Lindblad operators in the form of atomic densities ($\sum_s\psi_s^\dag(\bold{x})\psi_s(\bold{x})$). As the function $F(k_{eg}\bold{r})$ falls off on the length scale of laser wavelength, a spontaneous emission process will tend to localize a particle within this length scale, decohering the many-body state. $\mathcal{J}\rho$ together with $H_{\text{eff}}$ also preserves the trace of the density operator. 

We can obtain a multi-band Fermi-Hubbard model for the coherent part of the evolution in the master equation by expanding the field operators in a Wannier basis \cite{Kohn1959}, $\psi_s(\bold{x})=\displaystyle\sum_{n,i}w_{n}(\bold{x}-\bold{x}_i)c_{n,i,s}$, under the assumptions of nearest neighbor tunneling and local interaction in a deep lattice. Here, for the $i$-th site of the $n$-th Bloch band, $w_{n}(\bold{x}-\bold{x}_i)$ is the Wannier function and $c_{n,i,s}$ is the fermionic annihilation operator for spin $s$. In an isotropic 3D cubic lattice we get,

\begin{align}
\frac{d}{dt}\rho=-i[H,\rho]+\mathcal{L}\rho
\,,\end{align}
with the Fermi-Hubbard Hamiltonian,

\begin{align}
H = & -\displaystyle\sum_{n,<i,j>,s}J_{i,j,s}^{(n)}c^{(n)\dag}_{i,s}c^{(n)}_{j,s}+\displaystyle\sum_{n,i,s}\epsilon_{i,s}^{(n)}c^{(n)\dag}_{i,s}c^{(n)}_{i,s} \nonumber \\
 &+\displaystyle\sum_{i,k,l,m,n}U^{(k,l,m,n)}c^{(k)\dag}_{i,s}c^{(l)\dag}_{i,s'}c^{(m)}_{i,s'}c^{(n)}_{i,s}
\,.\end{align}
Here $J_{i,j,s}^{(n)}$ is the next neighbor tunneling rate corresponding to the kinetic energy, $U^{(k,l,m,n)}$ is onsite interaction energy coming mainly from collisional interaction with small modification from dipole interactions and $\epsilon_{i,s}^{(n)}$ is the onsite energy offset. The Lindblad term describing the scattering of laser photons:

\begin{align}
\mathcal{L}\rho=-\displaystyle\sum_{i,j,k,l,m,n,s,s'}\frac{\gamma^{k,l,m,n}_{i,j,s,s'}}{2}\left[c^{(k)\dag}_{i,s}c^{(l)}_{i,s},\left[c^{(m)\dag}_{j,s'}c^{(n)}_{j,s'},\rho\right]\right]
\,,\end{align}
and the matrix elements for different scattering processes: 

\begin{align}
\gamma^{k,l,m,n}_{i,j}=&\Gamma\int d^3xd^3y\frac{F(k_{eg}(\bold{x}-\bold{y}))}{4\Delta^2}\Omega^*(\bold{x})\Omega(\bold{y}) \nonumber \\
                       &w_k(\bold{x}-\bold{x}_i)w_l(\bold{x}-\bold{x}_i)w_m(\bold{x}-\bold{x}_j)w_n(\bold{x}-\bold{x}_j)
\,.\label{gamma}\end{align}

In Lamb-Dicke regime (i.e. Lamb-Dicke parameter, $\eta=k_La_0\ll1$ with $a_0$ as the extension of the Wannier functions in the lowest band), for a red detuned lattice spontaneous emissions dominantly return the atoms into the lowest Bloch Band \cite{Pichler2010} as the relative probability for the atom to return to the first excited band scales as $\eta^2$. Therefore we focus on the physics that arises from the treatment confined only to the lowest band and write down the corresponding master equation,

\begin{align}
\frac{d}{dt}\rho=-i[H_{FH},\rho]+\mathcal{L}_1\rho
\,.\end{align}

We now only have a single band Fermi-Hubbard Hamiltonian

\begin{align}
H_{FH}=-J\displaystyle\sum_{<i,j>,s}c^{\dag}_{i,s}c_{j,s}+U\displaystyle\sum_in_{i,\uparrow}n_{i,\downarrow}
\,,\label{FH}\end{align}
where we have omitted the band indices for the fermionic operators and the Liouvillian term is,

\begin{align}
\mathcal{L}_1\rho=\frac{\gamma}{2}\displaystyle\sum_{i}(2n_{i}\rho n_{i}-n_{i}n_{i}\rho-\rho n_{i}n_{i})
\,.\label{Lvln}\end{align}
Here $\gamma$ is the effective scattering rate obtained by keeping only the onsite elements in Eq.~\eqref{gamma} and the Lindblad operators $n_i$ are number operators at each site ($n_i=n_{i,\uparrow}+n_{i,\downarrow}=c^{\dag}_{i,\uparrow}c_{i,\uparrow}+c^{\dag}_{i,\downarrow}c_{i,\downarrow}$). It is clear at this point that the dissipative processes do not discriminate between the different spin orientations and can only decohere the many-body state by treating the lattice sites with different total particle numbers differently. In a system where particle numbers for each species are conserved individually, the term in the Hamiltonian corresponding to an energy offset is just a constant and thus can be neglected. Even though we have derived the master equation with two component systems in mind, the generalization to any number of internal states is straightforward and we can handle SU(N) magnetism with the same formalism. For simplicity, we will mainly focus on the physics of systems with two internal states for the rest of this article.

We now come back to the role of large detuning in avoiding direct spin decoherence when the transition frequencies differ for the two two level systems that represent the different spin states, i.e., $\omega_\uparrow \neq \omega_\downarrow$. We can modify the above derivation of the master equation at the expense of generating additional terms and look at the dynamics in this more general case. In the following we illustrate the effect for a single particle fixed in space at $\bold{x}_0$ having only two internal degrees of freedom. The corresponding master equation is given by

\begin{align}
\frac{d}{dt}\rho(t)=& -i\displaystyle\sum_s[\epsilon_sn_s,\rho] \nonumber \\
&+\displaystyle\sum_{s}\frac{\gamma_{s,s}}{2}\left(2n_s\rho n_s-n_sn_s\rho-\rho n_sn_s\right) \nonumber \\
&+\displaystyle\sum_{s\ne s'}\frac{\gamma_{s,s'}}{2}\left(2n_s\rho n_{s'}-n_sn_{s'}\rho-\rho n_sn_{s'}\right)
\,,
\label{singleparticle}
\end{align}
where  the spin dependent scattering rates are defined as follows
\begin{align}
\gamma_{s,s'}=\Gamma_s\int d^3x\frac{\Omega^*(\bold{x}_0)\Omega(\bold{x}_0)}{4\Delta_s\Delta_{s'}})|w_0(\bold{x}-\bold{x}_0)|^4
\,.\end{align}

The general solution for the atomic density matrix can be obtained analytically and is given by
\begin{equation*}
\rho(t) = \left(
\begin{array}{c c}
\rho_{\uparrow,\uparrow}(0) & \rho_{\uparrow,\downarrow}(0)e^{-(i\triangle\epsilon+\gamma_{\text{eff}})t} \\
\rho_{\downarrow,\uparrow}(0)e^{(i\triangle\epsilon-\gamma_{\text{eff}})t} & \rho_{\downarrow,\downarrow}(0)
\end{array} \right)
\,,\end{equation*}
where $\gamma_{\text{eff}}=\left(\gamma_{\uparrow,\uparrow}+\gamma_{\downarrow,\downarrow}-\gamma_{\uparrow,\downarrow}-\gamma_{\downarrow,\uparrow}\right)/2$ and $\triangle\epsilon=\epsilon_{\uparrow}-\epsilon_{\downarrow}$ and the associated decay rates $\Gamma_s$ can differ between spin states. 
The off-diagonal elements of the density matrix decay in magnitude exponentially with an effective rate $\gamma_{\text{eff}}$. This direct decoherence of the wave function is an effect of the spontaneous emission processes. Now in the limit of large detuning (i.e. $|\omega_{\uparrow}-\omega_{\downarrow}|/\Delta\to 0$) one can show, by taking a Taylor expansion of the function ${\Gamma_s/\Delta_s\Delta_{s'}}$ around any of the spin values, that the decay rate $\gamma_{\text{eff}}$ scales as $|\omega_{\uparrow}-\omega_{\downarrow}|/\Delta$. Therefore, for large detuning the master equation contains cross-terms of equal magnitude to the diagonal terms ($\gamma_{\text{eff}}\to 0$), and there is no direct decoherence in the system due to spontaneous emissions. On the technical level this means the Liouvillian part in Eq.~\eqref{singleparticle} reduces to a single particle and single-site version of Eq.~\eqref{Lvln}. This case of identical photon scattering is the standard case for fermionic atoms both from group-I and group-II in far detuned optical lattices.

\section{Decoherence in a double well}
\label{Sec:double}

\begin{figure}[t]
\centering
\includegraphics[width=0.5\textwidth]{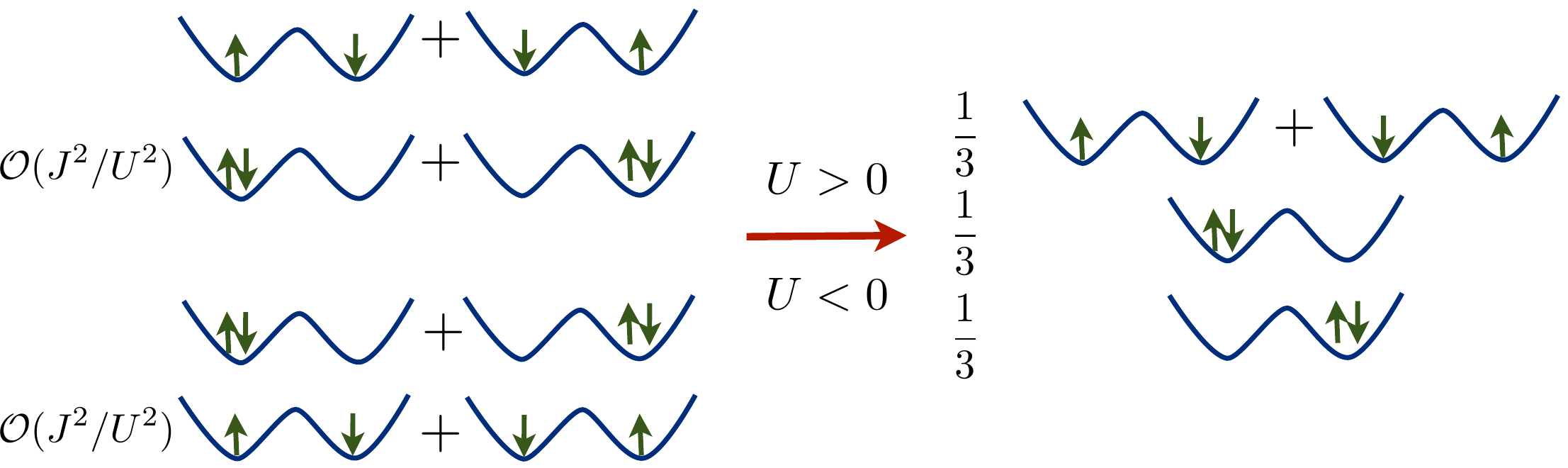}
\caption{Decoherence of fermions in a double-well. In the limit of strong interactions for $U>0$, the ground state of Fermi-Hubbard Hamiltonian with strong repulsive interactions is primarily a spin singlet (therefore symmetric spatially across the double-well). The population of doubly occupied sites is small ($\mathcal{O}(J^2/U^2)$). Now for $U<0$, the initial ground state is a coherent superposition of states with doubly occupied sites with $\mathcal{O}(J^2/U^2)$ population in the spin singlet state. Spontaneous emission events over a significant period of time lead to decoherence of virtual double-occupations, and populate states in which the final steady-state population is evenly distributed in the state with single occupancy and those with doubly occupied sites. \label{dw1}}
\end{figure}

We now proceed to study the effects of spontaneous emissions as described by the master equation derived in the previous section, focussing on the resulting many-body dynamics. We primarily take examples from strongly interacting regimes so that the spatial decoherence in the many body wave function due to localization of the spin particle following a spontaneous emission event is minimal \cite{Pichler2010}. We want to investigate the robustness of anti-ferromagnetic spin order of two species fermions in the repulsive case and of the correlation function of the composite bosons \cite{Greiner2003,Petrov2004,Takasu2003} formed in the case of strong attractive interactions. Before we present our results for larger lattice systems, we give an intuitive example discussing the decoherence in a double well.
For bosons, the dynamics of a related case is discussed in Ref.~\onlinecite{Poletti2013}. Here we particularly focus on the dynamics of the spin degree of freedom, which we treat first by considering the case of and initial ground state with $U>0$, $|U|\gg J$. We then return to the case of delocalised doublons for strong attractive interactions $U<0$.

\subsection{Repulsive interactions}

We consider an optical lattice chain with a length of two, containing one spin up particle and one spin down particle. Now, in the limit of strong interaction ($U\gg J$) the ground state would be a spin singlet with an admixture of states having both spins in one of the sites. It is instructive to work in a particular basis formed by combination of Fock states given by $|1\rangle=\left(|\uparrow,\downarrow\rangle+|\downarrow,\uparrow \rangle \right) /\sqrt{2}$, $|2\rangle =\left(|\uparrow,\downarrow\rangle-|\downarrow,\uparrow \rangle \right)/\sqrt{2}$, $|3\rangle=|\uparrow\downarrow,0\rangle$, $|4\rangle=|0,\uparrow\downarrow\rangle $. We first calculate the ground state of the two site Fermi-Hubbard Hamiltonian which is nearly a spin singlet with $\mathcal{O}(J^2/U^2)$ population in the manifold with double occupation at one of the sites (Fig.~\ref{dw1}). Evolving this initial state under the master equation shows that the wave function of the system decoheres due to spontaneous emission until it reaches a steady state ($\dot{\rho}=0$) where population is equally distributed in all three basis states that were populated at the initial time (Fig.~\ref{dw1}). The rate at which the spin correlation decays is proportional to that of the increase in the population of doubly occupied states (for a doubly occupied site $S^z=0$). We calculate this rate in perturbation theory in $J/U$ where the coherences between the manifolds are eliminated adiabatically to give the decay rate of the spin order. We begin by calculating the decay rate of population in state $|1\rangle$ which is being transferred to the doublet manifold spanned by $|3\rangle$ and $|4\rangle$. Now from the master equation,

\begin{figure}[t]
\centering
\includegraphics[width=0.45\textwidth]{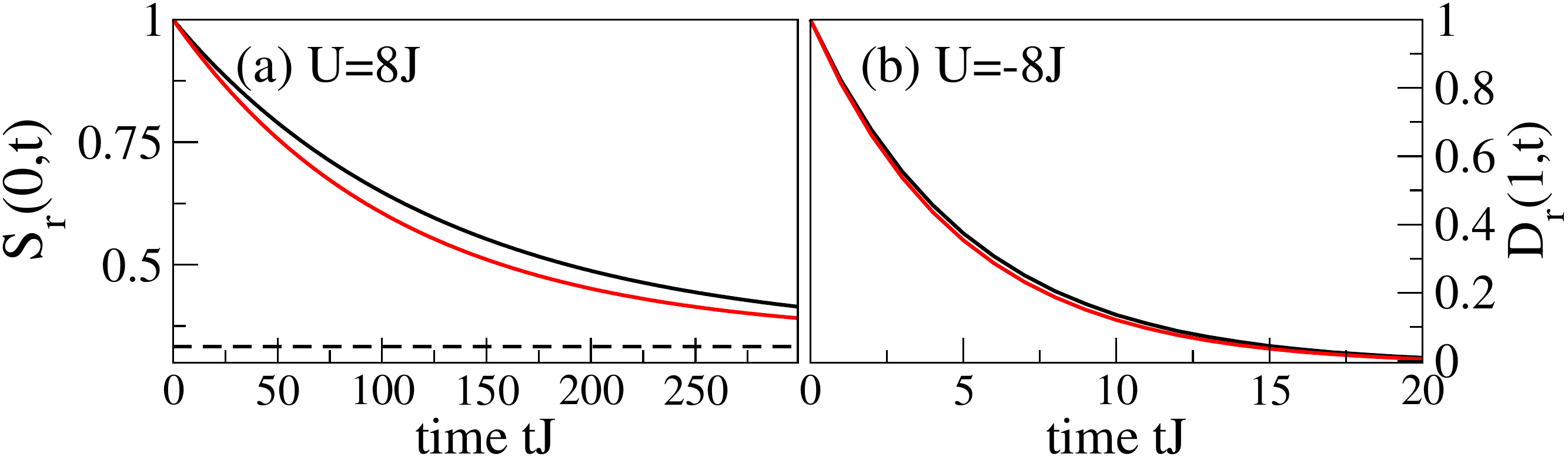}
\caption {Decoherent dynamics starting from the ground state of one particle of each spin species interacting strongly in a double well $(M=2)$ for $\gamma=0.1J$, computed in both, our perturbative approach (red dots) and exact diagonalization (solid line): (a) Decay of the rescaled spin correlations between the sites [Eq.~\eqref{SR}] for $U=8J$. The dashed line indicates the steady state expectation value. (b) Time evolution of rescaled doublon correlation [Eq.~\eqref{Dav}] for $U=-8J$ which vanishes in the final steady state. \label{dw2}}
\end{figure}

\begin{align}
\frac{d}{dt}\rho_{1,1} = -\displaystyle \sum_{k=3,4}2\sqrt{2}J\text{Re}\left(i\rho_{1,k}\right)
\,. \label{rho1}\end{align}
The coherences between state $|1\rangle$ and the doublet manifold obey

\begin{align}
\frac{d}{dt}\rho_{1,3} = i\sqrt{2}J \left( \rho_{3,3}+\rho_{4,3}-\rho_{1,1} \right)+(iU-\gamma)\rho_{1,3} 
\,,\end{align}
and $\rho_{1,4}$ follows an analogue equation. Now the coherence within the doublet manifold given by

\begin{align}
\frac{d}{dt}\rho_{3,4} = i\sqrt{2}J \left(\rho_{1,4}-\rho_{3,1}\right)-4\gamma\rho_{3,4}
\,.\end{align}
Now in the limit $U\gg J,\gamma$ we can eliminate the coherences between state $|1\rangle$ and the doublet manifold first and that leads us to

\begin{align}
\frac{d}{dt}\rho_{1,1} = & -\left(\frac{4J^2\gamma}{U^2+\gamma^2}\right)(2\rho_{1,1}-\rho_{3,3}-\rho_{4,4}) \nonumber \\
& +4J^2\text{Re}\left(\frac{\rho_{3,4}+\rho_{4,3}}{\gamma-iU}\right)
\,, \label{rho3}\end{align}
and

\begin{align}
\frac{d}{dt}\rho_{3,4}=-4\gamma\rho_{3,4}+\frac{4J^2\gamma}{U^2+\gamma^2}(\rho_{4,4}-\rho_{1,1}-\rho_{3,4})
\,. \label{rho4}\end{align}

Now we can eliminate the coherence in Eq.~\eqref{rho4} as it contains a term proportional to $\gamma$ whereas all the other terms are suppressed by a factor of $\mathcal{O}(J^2/U^2)$. We rewrite Eq.~\eqref{rho3} as

\begin{align}
\frac{d}{dt}\rho_{1,1}\approx-\left(\frac{4J^2\gamma}{U^2+\gamma^2}\right)(2\rho_{1,1}-\rho_{3,3}-\rho_{4,4})
\,, \label{rho5}\end{align}
which gives a decay rate proportional to $\beta=4J^2\gamma/(U^2+\gamma^2)$. The result obtained by evolving the master equation using exact diagonalization is in agreement to this anaytical value as illustrated in (Fig.~\ref{dw2}(a)). The spatial average of the spin correlation function is defined as, 

\begin{equation}
S(\Delta x,t)=\frac{1}{M}\sum_i \langle S^z_i(t)S^z_{i+\Delta x}(t)\rangle
\,.\label{Sav}
\end{equation}
Here, $S^z_i$ is the $z$ component of spin at lattice site $i$, defined as $S_i^z=\left(n_{i,\uparrow}-n_{i,\downarrow}\right)/2$. All spin components are equivalent due to the SU(2) symmetry of the lattice Hamiltonian and the dissipative terms. Therefore we focus on the $z$ component of spin and for plotting purposes we also consider the rescaled spatial average: 
\begin{equation}
S_r(\Delta x,t)=\frac{S(\Delta x,t)}{S(\Delta x,t=0)}
\,.\label{SR}
\end{equation}
The physical process giving rise to the observed decay can be outlined as follows: A spontaneous emission event does not differentiate between the different spin states, but as we saw before, what it effectively detects is the occupation number at the site involved as indicated by the Lindblad operators in Eq.~\eqref{Lvln}. In this sense, it distinguishes states with doubly-occupied sites from states with singly occupied sites, decohering virtual population of doubly occupied states. This drives the system away from the initial spin ordered ground state which has mostly singly occupied sites with very small doubly occupied population ($\mathcal{O}(J^2/U^2)$) and transfers population from states with singly occupied sites to states with doubly occupied sites. The resulting state is no longer an eigenstate and the Hamiltonian therefore starts redistributing population coherently whereas spontaneous emission events continue to disrupt rebuilding of coherence. This interplay between the Hamiltonian and the dissipative dynamics gives rise to the resulting decoherence and change in spin correlation. The rate of decoherence depends on the effective scattering rate $\gamma$ as well as on the relative population in the doublet manifold which grows proportionally with its initial value. This is the reason we have a term $\mathcal{O}(J^2/U^2)$ in the expression of $\beta$, and this reflects the ability of the Hamiltonian to populate doubly-occupied sites via tunneling in the presence of an energy gap. It also shows that the spin order decays much slower than the scattering rate and can be quite robust against spontaneous decay for strongly interacting systems.

The perturbative result for the decay rate obtained for the double well can be generalized for a chain of length $M$. By determining the initial relative population in the doublet manifold given by
\begin{equation}
 N_D=\frac{1}{N}\sum_i d_i^\dag d_i
\end{equation}
where $d_i=c_{i,\uparrow}c_{i,\downarrow}$ is the doublon annihilation operator, in degenerate perturbation theory \cite{Landau1977} in $J/
U$, where the Fermi-Hubbard Hamiltonian reduces to a Heisenberg Hamiltonian \cite{Cleveland1976} and systematic adiabatic elimination of coherences between the initial ground state in zeroth order and the doublet manifold gives exactly the same transfer rate. Furthermore, within the same perturbative approach we can directly relate our expression for the decay rate of the spin correlation, namely $\beta\propto \gamma N_D$. In Sec.~\ref{sec:repulsive} we will confirm the scaling predicted here for repulsive 1D systems up to $M=32$ lattice sites.

\subsection{Attractive interactions}
\label{sec:attractivedouble}
We now consider the case of attractive interactions ($U<0$), again for one atom of each spin in a double-well, where we observe markedly different dynamics for strong interactions. The ground state of the Hamiltonian now consists of states with double occupation, because the attractive interactions favour the formation of a dimer. The key physical property is that the dimer is delocalised over the two sites, i.e., the ground initial state is essentially a coherent superposition $(|3\rangle+|4\rangle)/\sqrt(2)$. Again, there is a small admixture of the singlet singly-occupied state, i.e., a population $\mathcal{O}(J^2/U^2)$ in $|1\rangle$. The final steady state of the master equation at long times is the same with equal population in all these three basis states. However, the initial dynamics towards the steady state begin by rapidly removing the coherence between $|3\rangle$ and $|4\rangle$, markedly changing the state when we consider the dynamics of dimers. We can calculate the decay rate of the doublon correlation functions, e.g., $d^{\dag}_1d_2$ in perturbation theory like before. Defining a spatially averaged and rescaled doublon correlation function analogous to the spin case
\begin{equation}
D_r(\Delta x,t)=\frac{1}{M}\frac{\sum_i\langle d^{\dag}_i(t)d_{i+\triangle x}(t) \rangle}{\langle d^{\dag}_i(t=0)d_{i+\triangle x}(t=0) \rangle}
\,,\label{Dav}
\end{equation}
we check the agreement between results obtained in exact diagonalization and perturbation theory (Fig.~\ref{dw2}(b)). The decay rate for the doublon correlations turns out to be $4\gamma$ and is approximately independent of the system size and filling factor, as we show up to first order in time-dependent perturbation theory in Sec.~\ref{sec:attractive}. In that section, we also discuss the enhancement factor, which arises from a combination of having two atoms in a given site, and also having superradiant enhancement because of the scattering of identical photons. An instructive way to check this enhancement factor is to look at the optical Bloch equations for a system of identical two-level atoms fixed on lattice sites and solve for an effective decay rate which is equivalent to calculating the rate of change in ground state population when the excited states can be adiabatically eliminated in the limit of large detuning ($\Delta$) of the driving laser field. For $N$ atoms the atomic density operator ($\rho_a$) obeys the following equation where the non-hermitian effective Hamiltonian $H'$, written in terms of Pauli matrices,

\begin{align}
\frac{d}{dt}\rho_a=-i[H',\rho_a]+\displaystyle\sum_{k,l}\Gamma_{kl}\sigma^-_{k}\rho_a \sigma^+_{l}
\,,\end{align}
with

\begin{align}
H'=\displaystyle\sum_{k=1}^N \left( -\Delta \sigma^z_{k} - \frac{\Omega_k}{2}(\sigma^+_{k}+\sigma^-_{k}) \right)-\frac{i}{2}\displaystyle\sum_{k,l}\Gamma_{kl}\sigma^+_{k} \sigma^-_{l}
\,.\end{align}
Here $\Omega_k$ is the Rabi frequency for the $k$-th atom (which we will take to be position independent) and $\Gamma_{kl}=\Gamma F(k_{eg}\bold{r}_{kl})$ where $\Gamma$ is just the spontaneous decay rate of the excited state of an atom and the function $F$ (Eq.~\ref{Ffn}) introduces a localising effect on the scattering element between $k$-th and $l$-th site (at a distance $r_{kl}$) on a scale set by the atomic transition wavelength $k_{eg}$. We can determine an effective decay rate in the ground state population which initially (when the system is the ground state $|\psi_0 \rangle$) is the rate of decrease in the norm for evolution under the effective Hamiltonian, namely,

\begin{align}
\Gamma_{\text{eff}}=&- \frac{1}{\delta t}[\langle \psi_0|(e^{iH'^{\dag}\delta t}e^{-iH'\delta t)}|\psi_0 \rangle] \nonumber \\
\approx& \langle \psi_0|\displaystyle\sum_{k,l}\Gamma_{kl}\sigma^-_{k} \sigma^+_{l}|\psi_0 \rangle
\,.\end{align}
In the single atom case, using second order time-dependent perturbation theory (the dipole coupling with the laser field is the perturbative part of the Hamiltonian), it is easy to calculate this effective decay rate $\Gamma_{\text{single}}=\Gamma\Omega^2/(4\Delta^2)$. Now for two atoms we look at two limiting cases. When the atoms are separated by a distance much larger than the atomic transition wavelength, the scattering elements turn into on-site terms ($\Gamma_{kl}\rightarrow\Gamma\delta_{k,l}$) and the effective rate is $2\Gamma_{\text{single}}$. This is what one would expect for the total decay rate of two independent entities. In the opposite case, where the atomic distance is much smaller than the transition wavelength, all the scattering elements become independent of the distance between the atoms ($\Gamma_{kl}\rightarrow\Gamma$) and we indeed obtain an effective decay rate of $4\Gamma_{\text{single}}$.

\section{Dynamics for many atoms}
\label{Sec:numerics}
 
In order to further quantify the impact of spontaneous emissions on many-body correlations we discuss the full many-body problem using approximate analytic and numerically exact solutions to the master equation derived in Sec.~\ref{Sec:master}. First we discuss the effects of spontaneous emissions on the momentum distribution of non-interacting fermions. Second we analyze the decoherence of antiferromagnetic spin order in the case of strong repulsive interactions, comparing time-dependent perturbation theory for $J/U\to0$ to numerical data. 
Depending on the system size we use either exact diagonalization or combine adaptive time-dependent DMRG with the quantum trajectory approach \cite{Daley2014}, to capture the decoherence and time-dependence of first order correlation functions in detail. While the first approach is exact, time-dependent DMRG is well established as a convenient means to model the real-time dynamics induced by stochastic processes in one-dimensional systems that remain close to equilibrium. 

Our main results on the repulsive case are that the spin correlation functions are robust on experimentally relevant timescales and that the spin-decoherence is governed by a single decay rate which is suppressed by the number of double occupancies in the initial state. In the limit of strong attractive interactions, both the perturbation theory approach and the numerical simulations unveil decay rates of the doublon correlation function enhanced by a factor of four, which can be understood as a consequence of superradiance \cite{Dicke1954,Gross1982,Lehmberg1970a}. While the results shown in this section are obtained from one-dimensional optical lattices, they are direct consequences of the derived master equation and general conclusions as the robust magnetic order and the impact of superradiance are therefore expected to carry over to both higher spin-degrees of freedom and higher dimensions. 

As a first general result we calculate the rate of energy increase induced by the spontaneous emissions for $N$ atoms. This can be obtained analytically from the master equation Eq.~\eqref{ME}, as was done for bosons in Ref.~\onlinecite{Pichler2010}, evaluating 
\begin{align}
\frac{d}{dt}H=\mathrm{Tr}({\mathcal{L}_1\rho H)}
\,.\end{align} 
The final result strongly resembles the result for bosons\cite{Pichler2010} and is not only independent of the interaction strength but also completely determined by single particle physics\cite{Gerbier2010}:

\begin{align} 
\frac{d}{dt}\langle H \rangle= \frac{\Gamma \Omega^2_0}{4\Delta^2}\frac{k^2}{2m}N  \quad \quad (\Omega=\Omega_0\cos{k_Lx}) \,.
\label{eq:rate}
\end{align}
However, as in the case of bosons, this result does not properly characterize the heating induced by spontaneous emissions as the energy increase predominantly results from excitations to higher bands which will in general not thermalize on experimental time-scales \cite{Pichler2010}. For Bosons this has been quantified in Ref.~\onlinecite{Schachenmayer2014}. Hence, even a qualitative analysis requires at least an analysis of first order correlation functions such as spin correlations, momentum distribution functions or the single particle density matrix. In the following we perform such an analysis, first for free Fermions, then for repulsive interactions and finally for attractive interactions.

\subsection{Free Fermions}
\label{sec:free}

The case of free fermions is another instructive example that can be dealt with exactly. We here focus on the time dependence of the momentum distribution for $N$ fermions on $M$ lattice sites
\begin{align}
n_k=\sum_{s}c^{\dag}_{k,s}c_{k,s},\text{ with } c_{k,s}=\sum_{l=1}^M\frac{1}{\sqrt{M}}e^{-ikl}c_{l,s}\,.
\end{align}
For $U=0$ the Hamiltonian is diagonal in momentum space and hence the time-evolution of $n_k(t)$ is solely given by the action of the dissipative part which results in
\begin{align}
&\frac{d}{dt}\langle n_k\rangle=\text{Tr}(n_k\mathcal{L}_1\rho)\nonumber\\
&=-\frac{\gamma}{2}\sum_{r,s}[n_r,[n_r,n_{k,s}]]\nonumber\\
&=-\gamma \langle n_k\rangle+\frac{N}{M}\gamma
\,. \end{align}
Therefore the steady state momentum distribution function $\langle n_k\rangle \rightarrow {N}/{M}$ for $t\to\infty$, i.e., the momentum distribution corresponding to all particles being localized in space by spontaneous emissions. The dynamics leading to this state occur gradually, as particles are spread throughout the Brillouin zone via spontaneous emissions. 

\subsection{Repulsive interactions}
\label{sec:repulsive}

We now move to the richer case of repulsively interacting fermions with magnetic ordering. In particular, we numerically study the decay of spin correlations in the 1D Hubbard model with repulsive interactions, for which understanding and characterizing the impact of different heating mechanisms on the characteristic correlation functions is an important step on the way to experimentally realize quantum magnetism. In one spatial dimension strong correlation effects give rise to interesting many-body effects such as the absence of long-range order, which can be utilized to benchmark experiments with ultra-cold atoms against exact solutions\cite{Essler2004}, and powerful numerical methods \cite{Daley2004}.
The one-dimensional Heisenberg model, one of the paradigm models of quantum magnetism\cite{Kolezhuk2004}, can be obtained from Fermi-Hubbard model using perturbation theory\cite{Cleveland1976} in $J/U$, which highlights the characteristic energy scale to observe quantum magnetism\cite{Meinert2013,Fukuhara2013,Simon2011}. Furthermore the antiferromagnetic correlations persist to finite $J/U$ and can be measured in experiments with ultra-cold fermions\cite{Meineke2012}. Here, we present the numerical data for the decay of the spatially averaged spin correlation functions defined in Eq. ~\eqref{Sav}. 
 
\begin{figure}[t]
\centering
\includegraphics[width=.5\textwidth]{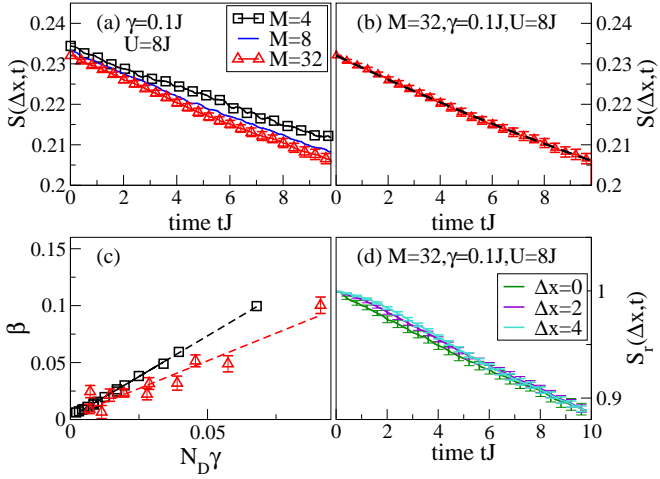}
\caption{Comparison of the decay of spin correlations averaged over the chain obtained from exact diagonalization using the EXPOKIT package \cite{Sidje1998} for $M=4,8$ and tDMRG with $D=128$ for $M=32$. (a) Decay of the on-site contribution $S(\Delta x=0,t)$ for different system sizes $M=4,8,32$ at $U=8J$ and $\gamma=0.1J$ using $500$ trajectories for $M=4,8$ and $250$ trajectories for $M=32$. (b) Example fit to the data shown in panel (a) for $M=32$. (c) The decay rates $\beta$ extracted from numerical fits as shown in panel (b) as a function of $N_D\gamma$. The dashed lines in (b) and (c) are linear fits to the data for different system sizes [M=4 (black squares) and M=32 (red triangles), which exhibit the scaling $\beta\sim N_D\gamma$ predicted by perturbation theory. Panel (d) shows the rescaled $S_r(\Delta x,t)$ for $M=32$, $U=8J$ and different $\Delta x$ which shows only a weak distance dependence, especially at larger times.
\label{DMRG plots}}
\end{figure}

Fig~\ref{DMRG plots} (a) shows the decay of the on-site contribution $S(\Delta x=0,t)$ at an interaction strength of $U=8J$ and $\gamma=0.1J$ for different system sizes $M=4,8,32$, a time-step of $dtJ=0.01$ and a DMRG matrix dimension of bond dimension of $8M$ to keep the discarded weight below $10^{-5}$ for the largest system at the largest time considered ($tJ=10$). We find that for the averaged quantities finite size effects are small for $M\geq 6$.
Fig. \ref{DMRG plots}(c) shows the main result, the decay rate $\beta$ extracted from a numerical fit of $a\cdot e^{-\beta t}+{\rm const.}$ [see Fig.~\ref{DMRG plots}(b) for an explicit example] to the curves as shown in Panel (a) for combinations of $U/J=4,6,8,10,12$ and $\gamma/J=0.2,0.1,0.05,0.025$ as a function of the effective decay rate obtained from perturbation theory, $N_D\gamma$. Within the error bars obtained from the fits, the decay rates obtained from  the numerical data scale linearly with respect to $N_D\gamma$  and the system size dependence is mainly given by system size effects of $N_D$. This corroborates our previously perturbative result, that the effective decay rate is suppressed as $U/J$ increases since it is proportional to the number of double occupancies in the initial state. Finally, Fig.~\ref{DMRG plots}(d) shows data for Eq.~\eqref{SR}, for distances $\Delta x=0,2,4$ at $M=32$, $U=8J$ and $\gamma=0.1J$. While we find that the alternating sign of $S(\Delta x,0)$ -- a necessary condition for antiferromagnetic correlations -- is preserved during the dynamics, rescaling the data according to Eq.~\eqref{SR} unveils that in addition the correlation function decays in a similar fashion independent of distance. 

To summarize, our numerical study of the decay of spin correlation functions for the repulsive Fermi Hubbard model undergoing spontaneous emissions shows that changes in antiferromagnetic correlations are inhibited because the rate is controlled by the number of double occupancies that can be formed. The energy gap plays an important role in suppressing the coherent processes that form virtually doubly-occupied sites, and leads to a suppression of the decay of magnetic correlations somewhat analogous to the inhibition of diffusion seen for Bosons in Refs.~\onlinecite{Poletti2012,Poletti2013}. The rate of doubly occupied sites is an experimentally controllable parameter \cite{Ronzheimer2013,Greif2013}, and the time-dependence of the spin correlations should be directly measurable in experiments, either using quantum gas microscopes \cite{Simon2011,Weitenberg2011}, or other techniques such as modulation spectroscopy or Bragg scattering to detect local or longer-range spin fluctuations \cite{Greif2013,Miyake2011}.
Note that this robustness shifts the typical rate of decay from $\gamma$ to $N_D \gamma \sim (J^2/U^2) \gamma$. This compares favourably with the energy scale $J^2/U$ of typical dynamics in this regime. Note that due to the suppression, the new dominant effect of spontaneous emissions for large enough $U$ will be transfer of particles to higher Bloch bands, on timescales given by $1/(\eta^2 \gamma)$. 

\subsection{Attractive interactions}
\label{sec:attractive}

This inhibition of the decay of spin correlation functions is in strong contrast to the effects we observe for attractive interactions, as we saw in the case of the double-well above. Here we analyse the characteristic correlation functions for many bosons with strong attractive interactions. Taking $U<0$ and focusing on strong interactions at moderate to low densities, we see that the ground state of the Fermi-Hubbard model consists of bound dimers that behave as composite bosons, and condense to allow condensation, and off-diagonal long-range order of dimers.

\begin{figure}[t]
\centering
\includegraphics[width=.5\textwidth]{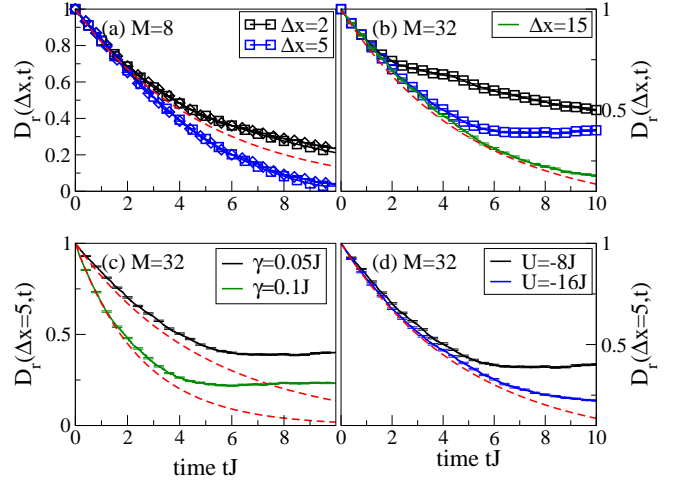}

\caption{Comparison of doublon correlation obtained numerically and in perturbation theory (dashed line) in the strong attractive interaction limit, averaged over chain and rescaled by the initial value. (a) Spatial dependence of doublon correlation for system size $M=8$ at $U=-8J$ and $\gamma=0.05J$.  We see that the quantum trajectory results (diamonds) from tDMRG with $D=64$ are in good agreement with the result obtained by doing exact diagonalization (squares) using the EXPOKIT package \cite{Sidje1998} and that perturbation theory does not take into account the rebuilding of correlations destroyed by spontaneous emissions and hence underestimates the decay at short distances, but overestimates the decay at large distances. Using same line symbols in panel (b) we show the quantum trajectory results for spatial dependence for $M=32$ at $U=-8J$ and $\gamma=0.05J$ qualitatively similar to $M=8$. (c) Effects of different decay rates for $M=32$ at $U=-8J$. (d) Dependence on interaction strength for $M=32$, $\gamma=0.05J$. The time for which our perturbation theory is reliable scales with $U$. (b) to (d) show tDMRG data using a bond dimension D=$128$ and the number of trajectories used in all of the calculations here is $528$.}
\label{fig:attractive}
\end{figure}

In the strongly interacting regime we expect pairs to predominantly form in real space and hence, for a sufficiently low density, the ground state of our lattice model has a large contribution of doubly occupied sites. To see this immediately we can again use degenerate perturbation theory in $J/U$, as was done in Ref.~\onlinecite{Petrosyan2007}, to find an effective Hamiltonian $H_D$ that describes the dynamics of bound pairs, 

\begin{equation}
H_D=\frac{2J^2}{U}\sum_{<i,j>}\left(d^{\dag}_id_j-n_i^{(D)}n_j^{(D)}\right)\,,
\end{equation}
which contains a doublon tunneling term as well as a nearest neighbour interaction term with $n_i^{(D)}=d^\dag_i d_i$ being the on-site number operator for doublons. This model favors pair formation on alternative sites as the system can decrease its energy via virtual tunneling of doublons ($U<0$). Since these pairs can be approximately treated as bosons and Pauli-exclusion prohibits multiple pairs, the perturbative Hamiltonian is the one of hardcore bosons with next-nearest neighbour interactions. At low densities we expect a superfluid of pairs for  the ground state with an algebraic decaying doublon correlation function (Eq.~\eqref{Dav}). Here we study the decay of those correlations during the dissipative dynamics. Given that the initial state is the ground state which is an eigenstate of the Hamiltonian, the first order of time-dependent perturbation theory is given solely by the action of the dissipative part on the initial state. We therefore calculate:

\begin{align}&
\frac{d}{dt}\langle d^{\dag}_id_j \rangle= \text{Tr}\left(d^{\dag}_id_j\mathcal{L}_1\rho\right) \nonumber \\
 &= \frac{\gamma}{2} \displaystyle\sum_{k} \langle 2n_{k}d^{\dag}_id_jn_{k}-n_{k}n_{k}d^{\dag}_id_j-d^{\dag}_id_j n_{k}n_{k} \rangle 
 \,,\end{align}
where

\begin{equation}
n_k=n_{k,\uparrow}+n_{k,\downarrow}
\,.\end{equation}
We first calculate

\begin{align}
& n_{k} d^{\dag}_id_j- d^{\dag}_id_j n_{k} \nonumber \\
&=-\left(c^{\dag}_{i,\uparrow}c^{\dag}_{i,\downarrow}c_{j,\downarrow}c_{k,\uparrow}+c^{\dag}_{i,\uparrow}c^{\dag}_{i,\downarrow}c_{k,\downarrow}c_{j,\uparrow}\right)\delta_{k,j}\nonumber\\
&\quad +\left(c^{\dag}_{k,\uparrow}c^{\dag}_{i,\downarrow}c_{j,\downarrow}c_{j,\uparrow}+c^{\dag}_{i,\uparrow}c^{\dag}_{k,\downarrow}c_{j,\downarrow}c_{j,\uparrow}\right)\delta_{k,i}\,,
\end{align}
and reinsert this identity to perform the sum over $k$ and obtain

\begin{align}
\frac{d}{dt}\langle d^{\dag}_id_j \rangle = \gamma \langle -2d^{\dag}_id_j-2d^{\dag}_id_j \rangle=-4\gamma \langle d^{\dag}_id_j \rangle
 \,.
 \label{dddecay}
\end{align}

Therefore the perturbative decay rate for the pair correlations is four times the scattering rate. Previous calculations for the Bose-Hubbard model show rates of decay for off-diagonal elements of the single particle density matrix for interacting bosons given \cite{Pichler2010} by the rate $\gamma$, so in our case the decay is thus four times larger. As noted above for a double-well, one factor of two arises as two particles form each dimer, whereas the other factor of two arises from the superradiant enhancement discussed in Sec.~\ref{sec:attractivedouble}. We expect our perturbative results to be valid only on a short time-scale set by the tunnelling rate, but as we show in Fig.~\ref{fig:attractive} this unexpected result persists through an numerical analysis for larger times and finite system sizes. Fig.~\ref{fig:attractive} (a) shows the spatial dependence of $D_r$ for system size $M=8$ at $U=-8J$ and $\gamma=0.05J$. We directly compare exact diagonalization (squares) and tDMRG data (diamonds) with our result from first order time dependent perturbation theory, a single exponential decay with the surprisingly high scattering rate of $4\gamma$. We find indeed that the data is well described by the perturbative result for up to $t\approx2/J$. For larger times the coherent dynamics neglected in the perturbation theory gives rise to two different types of behavior, depending on the spatial separation of the particles constituting the pair. For small $\Delta x$ we observe the rebuilding of the correlations between sites via tunneling after a spontaneous emission occurred. This rebuilding of correlations takes longer as the distance between the sites grows and therefore the deviation of the numerical data from the perturbative result becomes smaller. Fig.~\ref{fig:attractive} (b) shows a very similar result obtained for $M=32$ lattice sites using t-DMRG with quantum trajectories at $U=-8J$ and $\gamma=0.05J$. Similar to the repulsive site finite size effects in the averaged quantities are small, and the superradiance effect persists. In Fig.~\ref{fig:attractive} (c) we vary the decay rate $\gamma$ for $M=32$ at $U=-8J$, finding consistent behavior. Finally Fig.~\ref{fig:attractive}(d) probes the dependence on interaction strength for $M=32$, $\gamma=0.05J$ and we can see that the time for which perturbation theory is reliable scales with $U$.

As explained previously, this factor of four is due to the superradiance effect since the spatial separation of the dimers is much smaller than the wavelength of light. For the spontaneous emission from a doublon, the spatial separation is limited to the dimension of a single lattice site and hence much smaller than the wavelength of the light. In such a case the atoms interact with the light in a collective and coherent fashion\cite{Dicke1954,Gross1982}. This causes $N_l$ particles on site $l$ to spontaneously emit photons with a rate of $N_l^2\gamma$. 

This result is drastically different from exponential decay with rate $\gamma$ which we obtain for the single particle density matrix considering only singly occupied lattice sites. Note that superradiance \cite{Lehmberg1970a} does not depend on particle number statistics, hence the same enhanced decay rate is predicted for bosonic pair correlations, which is consistent with the bosonic version of the perturbative result stated by Eq.~\eqref{dddecay}. It is, however, important that we are in a regime where the photons scattered by atoms of different states are indistinguishable - this is the key origin of the superradiance in this case. Although this process changes the total scattering rate, it does not change the total rate of increase in energy. However, the change in the correlation functions should be directly measureable in ongoing experiments, with the off-diagonal correlations of dimers measurable by associating two particles on a specific lattice site to molecules, and measuring the momentum distribution of molecules.

\section{Summary and outlook}
\label{Sec:sum}

We have derived a microscopic master equation for the description of spontaneous emissions in two species of fermions in an optical lattice, specialising to the regime that is typical for atomic physics of cold atoms in optical lattices, where photons scattered from separate species are essentially indistinguishable. Because the scattered photons do not distinguish between spin states, but simply decohere superpositions of different local number states, magnetically ordered Mott Insulators are surprisingly robust, with the effects of spontaneous emissions within the lowest band being suppressed by a factor larger than the suppression of the dynamical timescale for strong repulsive interactions. The case of strong attractive interactions is markedly different, with a gas of dimers exhibiting a rate of decay for characteristic off-diagonal correlations that is not suppressed, and instead is further enhanced by superradiance.

Our predictions are directly accessible in ongoing experiments, and they provide a basis for characterising and controlling heating due to spontaneous emissions. For experimental realisations of magnetically ordered states, this is especially encouraging, as the dominant processes will be transfer of particles to higher bands.  This also implies that for many purposes, lattices that are blue-detuned ($\Delta>0$) rather than red-detuned ($\Delta<0$) therefore have no specific advantages, analogously to the case of the Mott Insulator for bosons. Moreover, our calculations generalize naturally to analogous states with SU(N) symmetry. The opposite is true for the case of bound dimers, where the rate of change of the correlations is equal to four times the scattering rate for a single atom. In this case, blue-detuned laser light, which could suppress spontaneous emissions by roughly an order of magnitude over red-detuned light, would be strongly advantageous, at least in terms of this heating mechanism.

On a theoretical level, these calculations are the starting point for many further interesting investigations into the many-body physics of these models. In particular, we now have a microscopic description within which to explore questions of thermalisation within the lattice. One of the first next steps will be to investigate thermalization between atoms the lowest band and atoms transferred to a higher band, followed by broader investigations of whether energy introduced is thermalised on experimental timescales \cite{Schachenmayer2014}. 

\begin{acknowledgements}

We thank Dan Boyanovsky, Randy Hulet, Wolfgang Ketterle, Ken O'Hara, Hannes Pichler,
Ulrich Schneider, Matthias Troyer and Peter Zoller for helpful and motivating
discussions. This work was supported by AFOSR grant
FA9550-13-1-0093. Computational resources were
provided by the Center for Simulation and Modeling at the University
of Pittsburgh. 
\end{acknowledgements}

\appendix

\section{Atomic physics considerations}
\label{sec:Atomic Physics}

Here we present a selection of examples of the necessary steps that lead to the results we obtained in Sec. I for the different transition rates including a complete table of the results. We start by solving the optical Bloch equations for two level and three level systems in presence of a far-detuned laser field and radiation bath. In the steady state the relative probability amplitude associated with the $i$-th excited level is $\Omega_i/2\Delta_i$, where $\Omega_i$ and $\Delta_i$ are the Rabi frequency and detuning for that level respectively. Next we connect this result with the hyperfine atomic structures of group-I and group-II atomic cases. Calculating the dipole matrix elements \cite{Metcalf1999} enables us to write down the excited state that a particular ground state would go to in presence of a laser with a particular polarization $q$. In general the corresponding excited state will be a superposition of different hyperfine states. The interference between the decay channels from these states give the resultant final decay rate to any of the ground states. In the following we give explicit examples for both atomic species.
 
For $^{171}\text{Yb}$ if we take the laser polarization to be along $\hat{\bold{z}}$-axis i.e. $\bold{E}=E\hat{\bold{e}}_0$, and apply it on $|g_{\uparrow}\rangle$ (Fig.~\ref{Yb}(a)), the atom, in the limit of large detuning, goes to an excited state which is a superposition of $P_{1/2}$ states with same $m_F$ (as polarization is linear) and we have

\begin{align}
|e\rangle \propto \left[\frac{1}{3\Delta}|e_{1,\uparrow}\rangle+\frac{\sqrt{2}}{3(\Delta+\delta_{\text{hfs}})}|e_{2,\uparrow}\rangle\right]
\,.\end{align}
The prefactors come from the different dipole matrix elements. Expansion of $|F,m_F\rangle$ basis into $|L,m_L;S,m_S;I,m_I\rangle$ basis reveals that for very large detuning $|e\rangle$ has the same nuclear spin as the starting ground state. Therefore to conserve the nuclear spin under experimental timescales the relative decay rate for a spin flip is suppressed and given by $\propto (\delta_{\text{hfs}}/\Delta)^2$. Below we give a table for different transition rates for different laser polarizations. There is an overall multiplicative factor $\sim (1/\Delta)^2$ for all the rates given. 
\begin{table}
\begin{center}
  \begin{tabular}{| c || c | c || c | c || c | c |}
    \hline
      & \multicolumn{2}{|c|| }{$q=0$} & \multicolumn{2}{|c|| }{$q=1$} & \multicolumn{2}{|c| }{$q=-1$} \\ \hline
      & $|g_{\uparrow}\rangle$ & $|g_{\downarrow}\rangle$ & $|g_{\uparrow}\rangle$ & $|g_{\downarrow}\rangle$ & $|g_{\uparrow}\rangle$ & $|g_{\downarrow}\rangle$ \\ \hline

    $|g_{\uparrow}\rangle$ & $1$ & $\left(\frac{\sqrt{2}}{3}\frac{\delta_{\text{hfs}}}{\Delta}\right)^2$ & $1$ & $0$ & $1$ & $\left(\frac{\sqrt{2}}{3}\frac{\delta_{\text{hfs}}}{\Delta}\right)^2$  \\ \hline

  \end{tabular}
  \caption{Matrix elements for the possible decay processes in Fig~\ref{Yb}}
\end{center} 
\end{table}

\begin{table*}
\begin{center}
  \begin{tabular}{| c | c | c | c | c | c | c |}
    \hline
      \multicolumn{7}{|c| }{$q=1$} \\ \hline
      & A & B & C & D & E & F \\ \hline

    B & $0$ & $1$ & $0$ & $0$ & $\left(\frac{2\sqrt{2}}{9}(\beta_3)\right)^2$ & $\left(\frac{2}{3\sqrt{3}}(\beta_3)\right)^2$  \\ \hline

    E & $0$ & $\left(\frac{2\sqrt{2}}{9}(\beta_3)\right)^2$ & $0$ & $0$ & $1$ & $\left(\frac{\sqrt{2}}{3\sqrt{3}}\beta_3\right)^2$ \\ \hline

    F & $0$ & $0$ & $0$ & $0$ & $0$ & $1$ \\ \hline   

  \end{tabular}
\end{center}

\begin{center}
  \begin{tabular}{| c | c | c | c | c | c | c |}
    \hline
      \multicolumn{7}{|c| }{$q=0$} \\ \hline
      & A & B & C & D & E & F \\ \hline

    B & $\left(\frac{\sqrt{2}}{9}\beta_3\right)^2$ & $1$ & $0$ & $\left(\frac{2}{9}\beta_3\right)^2$ & $\left(\frac{2\sqrt{2}}{3}(\beta_1-\beta_2)\right)^2$ & $\left(\frac{2}{3\sqrt{3}}\beta_3\right)^2$  \\ \hline

    E & $\left(\frac{2}{9}\beta_3\right)^2$ & $\left(\frac{2\sqrt{2}}{27}(\beta_1-\beta_2)\right)^2$ & $0$ & $\left(\frac{2\sqrt{2}}{9}\beta_3\right)^2$ & $1$ & $\left(\frac{\sqrt{2}}{3\sqrt{3}}\beta_3\right)^2$ \\ \hline

    F & $0$ & $\left(\frac{2}{9\sqrt{3}}\beta_3\right)^2$ & $0$ & $0$ & $\left(\frac{\sqrt{2}}{9\sqrt{3}}\beta_3\right)^2$ & $1$ \\ \hline   

  \end{tabular}
\end{center}

\begin{center}
  \begin{tabular}{| c | c | c | c | c | c | c |}
    \hline
      \multicolumn{7}{|c|}{$q=-1$} \\ \hline
      & A & B & C & D & E & F \\ \hline
  
      B & $\left(\frac{\sqrt{2}}{9}\beta_3\right)^2$ & $1$ & $\left(\frac{2\sqrt{2}}{9\sqrt{3}}(\beta_1-\beta_2\right)^2$ & $\left(\frac{2}{9}\beta_3\right)^2$ & $\left(\frac{2\sqrt{2}}{9}\beta_3\right)^2$ & $0$ \\ \hline

      E & $\left(\frac{2}{9}\beta_3\right)^2$ & $\left(\frac{2\sqrt{2}}{9}\beta_3\right)^2$ & $\left(\frac{4}{9\sqrt{3}}(\beta_1-\beta_2)\right)^2$ & $\left(\frac{2\sqrt{2}}{9}\beta_3\right)^2$ & $1$ & $0$ \\ \hline

      F & $\left(\frac{2\sqrt{2}}{9\sqrt{3}}(\beta_1-\beta_2)\right)^2$ & $\left(\frac{2}{3\sqrt{3}}\beta_3\right)^2$ & $0$ & $\left(\frac{4}{3\sqrt{3}}\left(\beta_1-\frac{4}{5}\beta_2\right)\right)^2$ & $\left(\frac{\sqrt{2}}{3\sqrt{3}}\beta_3\right)^2$ & $1$ \\ \hline

\end{tabular}
\end{center}
\caption{Matrix elements for the possible decay processes in Fig~\ref{Li}}
\end{table*}
Now for the $^6\text{Li}$ atom starting at $|g_D\rangle$ (Fig.~\ref{Li}) with a linearly polarized laser $(q=0)$, the excited state would again be a superposition of states in $P$ sub levels with same nuclear spin component:

\begin{align}
|e\rangle \propto & -\frac{2\sqrt{2}}{9\Delta_1}|e_1\rangle-\frac{1}{9\Delta_1}|e_4\rangle-\frac{1}{9\Delta_2}|e_7\rangle-\frac{2}{9\sqrt{5}\Delta_2}|e_{10}\rangle \nonumber \\ 
&+\frac{1}{\sqrt{5}\Delta_2}|e_{15}\rangle
\, , \end{align}
with $\Delta_1=\Delta$ and $\Delta_2=\Delta_1+\delta_{\text{fs}}$. There are also possibilities of two different types of spin flips here. Considering a decay towards $|g_A\rangle$ we see that this state is orthogonal to $|g_D\rangle$ in terms of the combination of electron and nuclear spins, namely, in the $|L,m_L;S,m_S;I,m_I\rangle$,

\begin{align}
|g_D\rangle\propto |0,0\rangle\otimes\left(|\frac{1}{2},-\frac{1}{2};1,0\rangle+\alpha|\frac{1}{2},\frac{1}{2};1,-1\rangle\right)
\,.\end{align}
and,

\begin{align}
|g_A\rangle\propto |0,0\rangle\otimes\left(\alpha|\frac{1}{2},-\frac{1}{2};1,0\rangle-|\frac{1}{2},\frac{1}{2};1,-1\rangle\right)
\,,\end{align}
whereas the spin part of the excited state inside each $P$ sublevel looks like that of $|g_D\rangle$. Therefore the contributions from the hyperfine states in each $P$ sublevel cancel each other given the detuning is large compared to hyperfine structure splitting. A different mechanism of cancellation occurs if we consider a spin flip resulting in the state 
\begin{align}
|g_E\rangle\propto |0,0\rangle\otimes\left(|\frac{1}{2},\frac{1}{2};1,0\rangle+\beta|\frac{1}{2},-\frac{1}{2};1,1\rangle\right)
\,,\end{align}
This is a matrix element for a transition to a state with different electron spin than $|g_D\rangle$. In this case the paths via the two sublevels cancel each other and we obtain a suppression of spin flip as the detuning is large compared to the fine structure splitting. Here also we give a table for all the different transition rates for different laser polarizations. There is an overall multiplicative factor $\sim (1/\Delta)^2$ for all the rates given and we define $\beta_1=\delta_{\text{hfs},P_{1/2}}/\Delta$, $\beta_2=\delta_{\text{hfs},P_{3/2}}/\Delta$ and $\beta_3=\delta_{\text{fs}}/\Delta$. The starting states are chosen from the states with positive $z$-component of total angular momentum in the ground state manifold, as we can perform the same calculations for the other half symmetrically.

\begin{figure}[t]
\centering
\includegraphics[width=0.5\textwidth]{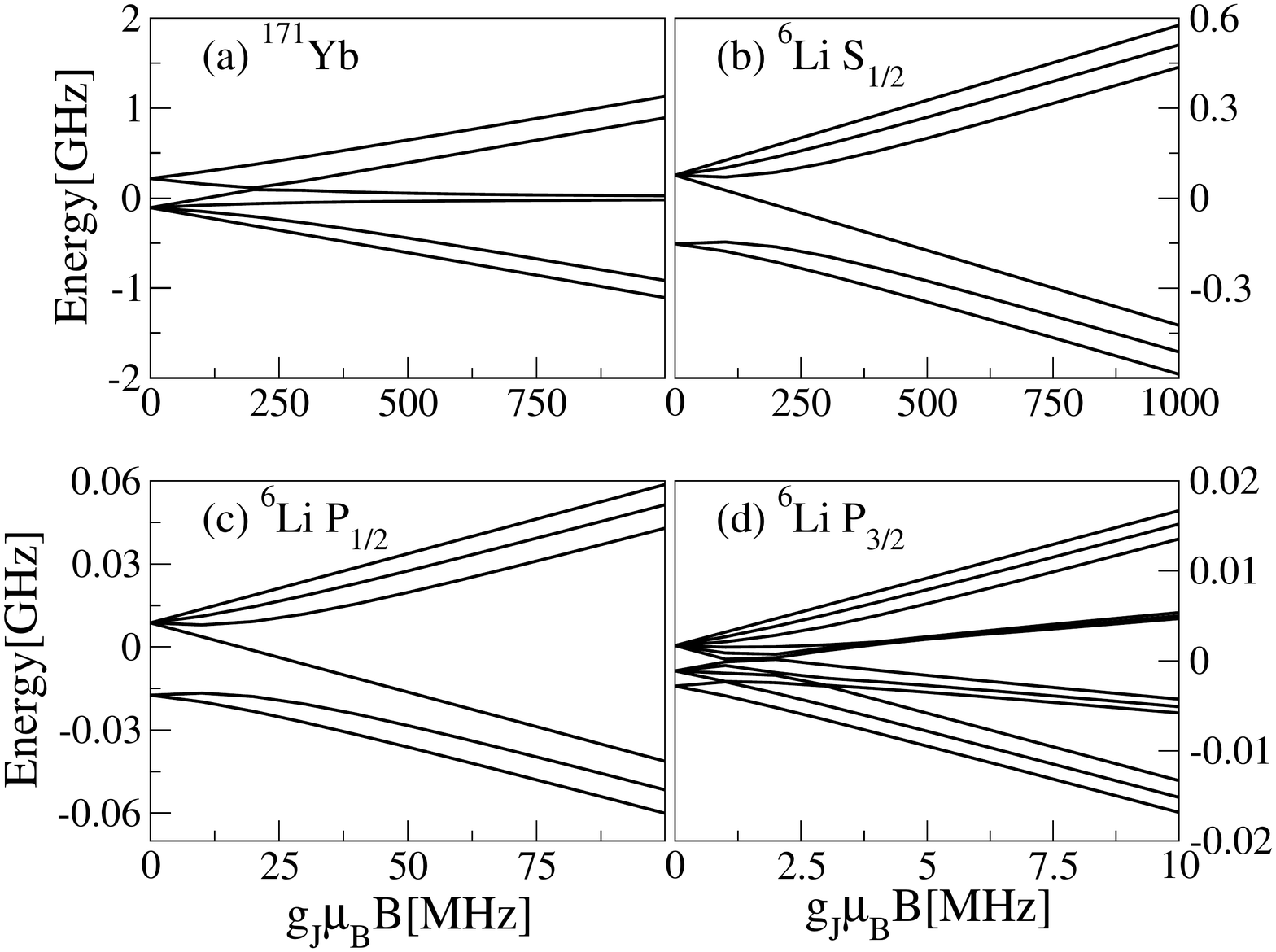}
\caption{Zeeman diagram of the different hyperfine levels shows the lifting of degeneracy obtained by numerical diagonalization for qualitative values of magnetic field. The electron and nuclear spins get decoupled in high enough B field. (a) $P_{1}$ sublevel in $^{171}\text{Yb}$ (the $S$ sublevel is already decoupled). (b) Splitting of the $S_{1/2}$ sublevel in $^{6}\text{Li}$. (c) $P_{1/2}$ sublevel in $^{6}\text{Li}$ needs weaker B field to get decoupled than the previous case as the hyperfine coupling is weaker. (d) Even smaller B field is needed for the even more weakly hyperfine-coupled $P_{3/2}$ sublevel in $^{6}\text{Li}$. 
}
\label{fig:fields}
\end{figure}

The field dependence of the energy levels used is showcased in Fig.~\ref{fig:fields}. We find that the probability of spin flip processes stay negligible across the whole range of field strengths, provided that the detuning is still much larger than the hyperfine coupling strength. Remarkably, we find that these rates do not change substantially as a function of the magnetic field. Moreover, these rates are so low in this context that we expect spin-flips to be dominated by other physical processes, such as transitions due to blackbody radiation. 


\section{Many Body Master Equation}
\label{sec:Master Equation}

\subsection{N-atom Optical Bloch Equation}

Here we present the detailed derivation of the master equation (Eq.~\eqref{ME}) for identical photon scattering. The dipole moment of the transition for each system is $\bold{d}_{eg}$. In our treatment we can take these moments to be spin independent as their value only depend on the radial part of the atomic wavefunction which is same for both the spin states. The interaction of the system with the electromagnetic field is treated under the dipole approximation. The driving laser field is described classically and is given by:
\begin{align}  
\bold{E}_{\text{cl}}(\bold{x},t)=\bold{E}_{\text{cl}}^{(+)}(\bold{x},t)e^{i\omega_Lt}+\bold{E}_{\text{cl}}^{(-)}(\bold{x},t)e^{-i\omega_Lt}   
\,.\end{align} 
The system is coupled to the quantized radiation field,
\begin{align}  
\bold{E}(\bold{x},t)&=\bold{E}^{(+)}(\bold{x},t)+\bold{E}^{(-)}(\bold{x},t) \nonumber \\ 
&=\sum_{\bold{k},\lambda}\left(\varepsilon_k\bold{e}_{\lambda,\bold{k}}e^{i\bold{k}\cdot\bold{x}}b_{\lambda,\bold{k}}+\text{h.c.}\right) 
\,,\end{align}  
where
\begin{align} 
\varepsilon_k=i\sqrt{\frac{\omega_k}{2\epsilon_o(2\pi)^3}} 
\,.\end{align} 
The Hamiltonian describing the evolution of the system is given by ($\hbar\equiv 1$):
\begin{align}  
H=H_0+H_I+H_F 
\,.\end{align} 
Here, the first term describes the atomic Hamiltonian,
\begin{align}
  H_0 = & \int d^3x\psi^\dag(\bold{x},t)\Big(\frac{-\nabla^2}{2m}+\omega_{eg}\sum_{s=\uparrow,\downarrow}|e_s\rangle \langle e_s|\Big)\psi(\bold{x},t)  .
\,.\end{align}
The second term is the dipole coupling between the atoms and the electric field,
\begin{align}
 &  H_I \nonumber \\
 & = -\sum_{s=\uparrow,\downarrow}\int d^3x\psi^\dag(\bold{x},t)(\sigma_{+,s}\bold{d}_{eg}+\text{h.c.})\cdot\bold{E}(\bold{x},t)\psi(\bold{x},t)  \nonumber \\ 
 & = -\sum_{s=\uparrow,\downarrow}\int d^3x\psi^\dag(\bold{x},t)(\sigma_{+,s}+\sigma_{-,s})\bold{d}_{eg}\cdot\bold{E}(\bold{x},t)\psi(\bold{x},t)
\,.\end{align}
The external radiation field Hamiltonian is given by,

\begin{align} 
H_F=\sum_{\bold{k},\lambda}\omega_kb^{\dag}_{\lambda,\bold{k}}b_{\lambda,\bold{k}}
\,.\end{align} 

Here we have fermionic field operator $\psi(\bold{x},t)$ (see Fig.~\ref{Yb} (b)) 
 
\begin{equation*}
\psi(\bold{x},t) = \left(
\begin{array}{c}
\psi_{e,\uparrow}(\bold{x},t) \\
\psi_{g,\uparrow}(\bold{x},t) \\
\psi_{e,\downarrow}(\bold{x},t) \\
\psi_{g,\downarrow}(\bold{x},t)
\end{array} \right)
\,,\end{equation*}
and raising and lowering operators for the different spins, $\sigma_{\pm,s}$.

The bosonic operators $b_{\lambda,\bold{k}}$ ($b^{\dag}_{\lambda,\bold{k}}$) annihilate (create) a photon in the mode ($\bold{k},\lambda$). From this Hamiltonian we get Heisenberg equations of motions for these operators of the quantized radiation field:

\begin{align}
\frac{d}{dt} & b_{\lambda,\bold{k}} =-i\omega_kb_{\lambda,\bold{k}} \nonumber \\
& +ig^*_{\lambda,\bold{k}}\sum_{s=\uparrow,\downarrow}\int d^3x\psi^\dag(\bold{x},t)(\sigma_{+,s}+\sigma_{-,s})\psi(\bold{x},t)e^{-i\bold{k}\cdot\bold{x}}  
\,,\end{align}
with $g_{\lambda,\bold{k}}=\varepsilon_k\bold{e}_{\lambda,\bold{k}}\cdot\bold{d}_{eg}$.

To solve this we make use of the Born-Markov approximation that emerges from the fact that the timescale set by the optical frequency is much faster than the other timescales in the problem, namely detuning, Rabi frequency and decay rate of the excited states. Under this approximation we can write,

\begin{align}
& \psi^\dag(\bold{x'},\tau)\sigma_{\pm,s}(\tau)\psi(\bold{x'},\tau) \nonumber\\
&\longrightarrow\psi^\dag(\bold{x'},t)\sigma_{\pm,s}(t)\psi(\bold{x'},t)e^{\mp i\omega_{eg}(t-\tau)}
\,.\end{align} 
Using this and defining

\begin{align}
\Omega^{\pm}(\bold{r})=\frac{\Gamma}{2\pi k^3_o}\mathcal{P}\int dk\frac{k^3F(k\bold{r})}{k\pm k_{eg}}
\,,\end{align}
with $\Gamma$ being the Wigner-Weisskopf spontaneous decay rate, we find a quantum-Langevin form of equation of motion for an operator $a$ acting only on the atomic degrees of freedom, by making use of rotating wave approximation and neglecting terms of the form $\psi^\dag(\bold{x},t)\sigma_{\pm,s}(t)\psi(\bold{x},t)\psi^\dag(\bold{y},t)\sigma_{\pm,s}(t)\psi(\bold{y},t)$ as they oscillate at double the optical frequency. We end up with

\begin{align}
\frac{d}{dt}a=i[H_0+H_{\text{cl}}+H_{\text{dip}},a]+\mathcal{L}a
\,,\end{align}
where the coupling Hamiltonian is

\begin{align}
 H_{\text{cl}}=-\sum_{s=\uparrow,\downarrow}\int d^3x\psi^\dag(\bold{x},t&)(\sigma_{+,s}\bold{d}_{eg}\cdot\bold{E}^+_{\text{cl}}(\bold{x},t) \nonumber \\
&+\sigma_{-,s}\bold{d}_{eg}\cdot\bold{E}^-_{\text{cl}}(\bold{x},t))\bold\psi(\bold{x},t) 
\,.\end{align}
The term describing dipole interactions is 

\begin{align}
& H_{\text{dip}}=\Gamma \int d^3xd^3yG(k_{eg}\bold{r})\,\times\nonumber \\ &\Big(\sum_{s=\uparrow,\downarrow}\psi^\dag(\bold{y},t)\sigma_{-,s}(t)\psi(\bold{y},t)\Big) 
\times\nonumber \\ &\Big(\sum_{s'=\uparrow,\downarrow}\psi^\dag(\bold{x},t)\sigma_{+,s'}(t)\psi(\bold{x},t)\Big)
\,.\end{align}
The term describing the dissipative dynamics is 

\begin{align}
&\mathcal{L}a =\int d^3xd^3y\frac{\Gamma}{2}F(k_{eg}\bold{r})\,\times \nonumber \\
&\sum_{s,s'=\uparrow,\downarrow}\Bigg\{ 2\Big(\psi^\dag_{e,s}(\bold{x},t)\psi_{g,s}(\bold{x},t)\Big)a\Big(\psi^\dag_{g,s'}(\bold{y},t)\psi_{e,s'}(\bold{y},t)\Big) \nonumber \\
 &  -\Big(\psi^\dag_{e,s}(\bold{x},t)\psi_{g,s}(\bold{x},t)\Big)\Big(\psi^\dag_{g,s'}(\bold{y},t)\psi_{e,s'}(\bold{y},t)\Big)a \nonumber \\
&-a\Big(\psi^\dag_{e,s}(\bold{x},t)\psi_{g,s}(\bold{x},t)\Big)\Big(\psi^\dag_{g,s'}(\bold{y},t)\psi_{e,s'}(\bold{y},t)\Big) \Bigg\}
\,.\end{align}
The diagonal term is an (infinite) Lamb shift that has been absorbed into a redefinition of the transition frequency. 

For coherent input states, corresponding to the classical laser field, we can equivalently write the master equation for atomic density operator $\rho$ (in rotating frame with laser frequency $\omega_L$),

\begin{align}
\frac{d}{dt}\rho=-i[H_0+H_{\text{cl}}+H_{\text{dip}},\rho]+\mathcal{L}\rho
\,.\end{align} \\

\subsection{Adiabatic elimination}

In the limit of large detuning the population in the excited states is negligible compared to that in the ground state and we can write down the master equation solely in terms of ground state filed operators. The precise conditions needed for this requires the detuning $\Delta$ to be much larger than Rabi frequency ($\Omega(\bold{x})=2\bold{E}_{\text{cl}}(\bold{x})\cdot\bold{d}_{eg})$, spontaneous decay rate $\Gamma$, decay rate times the number of particles in a volume $\lambda_L^3$ ($\Gamma\langle\psi^\dag(\bold{x})\psi(\bold{x})\rangle\lambda_L^3$), particle kinetic energy and dipole-dipole interaction between the particles. \\

Under these conditions we can solve the Heisenberg equation of motion for $\psi^\dag_{g,+}(\bold{z})\psi_{e,+}(\bold{z})$ and obtain 

\begin{align}
\psi^\dag_{g,+}(\bold{z})\psi_{e,+}(\bold{z})\approx & -\frac{\Omega(\bold{z})}{2\Delta}e^{-i\omega_Lt}\psi^\dag_{g,+}(\bold{z})\psi_{g,+}(\bold{z})
\,.\end{align}
Treating the other terms similarly we find a master equation for the atoms in ground state. From here onwards we will follow the convention in the main text, by omitting the index $g$ in field operators. We can write the master equation with an effective Hamiltonian

\begin{align}
\frac{d}{dt}\rho=-i\Big(H_{\text{eff}}\rho-\rho H_{\text{eff}}^{\dag}\Big)+\mathcal{J}\rho
\,,\end{align}
with non-hermitian effective Hamiltonian
\begin{align}
H_{\text{eff}}=H_0+H^{\text{light}}_{\text{eff}}
\,.\end{align}
The first term, $H_0$ describes the motion of the single atoms in the optical lattice,
\begin{align}
\sum_s\int d^3x\psi_s^\dag(\bold{x})\Big(\frac{-\nabla^2}{2m}+\frac{|\Omega(\bold{x})|^2}{4\Delta}\Big)\psi_s(\bold{x}) 
\,.\end{align}
The radiative part describes the couplings between the atoms and the vacuum modes of the electromagnetic field,
\begin{align}
&H^{\text{light}}_{\text{eff}}\nonumber \\
&= \sum_{s,s'}\Gamma\int G(k_{eg}\bold{r})\frac{\Omega(\bold{y})\Omega^*(\bold{x})}{4\Delta^2}\psi_s^\dag(\bold{x})\psi_{s'}^\dag(\bold{y})\psi_{s'}(\bold{y})\psi_s(\bold{x}) \nonumber\\    
                   & -i\frac{\Gamma}{2}\sum_s\int d^3x\frac{|\Omega(\bold{x})|^2}{4\Delta^2})\psi_s^\dag(\bold{x})\psi_s(\bold{x})\nonumber\\    
                   & -i\frac{\Gamma}{2}\sum_{s,s'}\int \frac{\Omega(\bold{y})\Omega^*(\bold{x})}{4\Delta^2} F(k_{eg}\bold{r})\psi_s^\dag(\bold{x})\psi_{s'}^\dag(\bold{y})\psi_{s'}(\bold{y})\psi_s(\bold{x})
\,.\end{align}
The recycling term is
\begin{align}
\mathcal{J}\rho=\Gamma\int d^3xd^3y&\frac{\Omega(\bold{y})\Omega^*(\bold{x})}{4\Delta^2} F(k_{eg}\bold{r})\nonumber \\
&\Big(\sum_s\psi_s^\dag(\bold{y})\psi_s(\bold{y})\Big)\rho\Big(\sum_s\psi_s^\dag(\bold{x})\psi_s(\bold{x})\Big)
\,.\end{align}
The master equation describes the motions of the atoms in a light field assuming they are well separated compared to the range of the interaction potential. We need to add a two body collisional interaction term $H^{\text{int}}_{\text{eff}}$ to the effective Hamiltonian:
\begin{align}
H^{\text{int}}_{\text{eff}}=\int d^3xg(x)\psi_{\uparrow}^\dag(\bold{x})\psi_{\downarrow}^\dag(\bold{x})\psi_{\downarrow}(\bold{x})\psi_{\uparrow}(\bold{x})
\,,\end{align}
where the true potential has been modeled by a pseudo-potential in which is a contact potential with the scattering length contained in $g(\bold{x})$. The spatial dependence is due to the laser intensity driven modification of the scattering length near an optical Feshbach resonance.

\bibliographystyle{apsrev}
\bibliography{fermion_bib}

\end{document}